\documentclass[aps,11pt]{revtex4}
\usepackage{dcolumn}
\usepackage{graphicx}
\usepackage{amsmath}
\usepackage{amsfonts}
\usepackage{amssymb}
\usepackage{psfrag}
\usepackage{wrapfig}
\usepackage{subfigure}
\usepackage{makeidx}
\usepackage{bm}
\usepackage{epsf}
\usepackage{hyperref}
\usepackage{color}
\usepackage{cases}
\usepackage{multirow}
\usepackage{float} 
\usepackage{array}

\makeatletter

\usepackage{booktabs} 
\usepackage{makecell} 
\usepackage{multirow}
\usepackage{bm}

\makeatletter
\newcommand{\eq}[1]{Eq.(\ref{#1})}

\newcommand{\Rmnum}[1]{\uppercase\expandafter{\romannumeral #1}}

\newcolumntype{M}[1]{>{\centering\arraybackslash}m{#1}}

\newcolumntype{M}[1]{>{\centering\arraybackslash$}m{#1}<{$}}  

\usepackage{amsmath} 

\usepackage{booktabs}

\begin{document} 

\title{Polar perturbations of dilaton-Euler-Heisenberg  black holes}

\author{Sheng-Yuan Li$^{1}$\footnote{shengyuanli77@outlook.com}, Yun Soo Myung$^{2}$\footnote{ysmyung@inje.ac.kr}, Ming Zhang$^{3}$\footnote{mingzhang0807@126.com}, \\ Xufen Zhang$^{1}$\footnote{xfzhangyzu@126.com} and De-Cheng Zou$^{4}$\footnote{Corresponding author:dczou@jxnu.edu.cn}}

\address{
$^{1}$Center for Gravitation and Cosmology, College of Physical Science and Technology, Yangzhou University, Yangzhou 225009, China\\
$^{2}$Center for Quantum Spacetime, Sogang University, Seoul 04107, Republic of  Korea\\
$^{3}$Faculty of Science, Xihang University, Xi'an 710077 China\\
$^{4}$College of Physics and Communication Electronics, Jiangxi Normal University, Nanchang 330022, China}
\date{\today}

\begin{abstract}
\indent

We investigate the quasinormal modes of polar metric-dilaton  perturbations around the dilaton-Euler-Heisenberg (dEH)  black holes with dilaton hair. The dEH black holes are obtained  from the  Einstein–Maxwell–dilaton theory with two dilaton coupling parameters ($\alpha,\beta$)  to the nonlinear  Euler–Heisenberg term. 
We compute the quasinormal mode spectra by making  use of two numerical techniques: direct integration and  matrix values continued fraction methods. 
An excellent agreement is found  between two approaches, confirming  the robustness of our computation. We present the fundamental quasinormal frequencies for both gravitational and dilaton modes and analyze their dependence on the magnetic charge ($Q_m$), angular momentum quantum number ($l$), and coupling parameter ($\epsilon=\alpha-\beta$).  All negative imaginary  quasinormal frequencies for polar metric-dilaton perturbations    imply  that the dEH black hole  with dilaton hair is stable against  dilaton with $l=0,1,2,3$ and gravitational modes with $l=2,3$.   Also, our results reveal distinct qualitative behaviors between  $\epsilon=1$ and $\epsilon= -1 $, particularly in the damping rates near the extremality.
\end{abstract}


\maketitle

\section{Introduction}
\label{intro}

As a nonlinear extension of quantum electrodynamics (QED), the Euler-Heisenberg (EH) Lagrangian formulated in 1936 \cite{Heisenberg:1936nmg} has  provided a classical approximation superior to the  Maxwell theory under strong-field condition where vacuum polarization becomes significant. Within this framework,  the vacuum is regarded as a dynamically polarizable medium  with polarization/magnetization arising from virtual charge clouds around real charges/currents \cite{Obukhov:2002xa}.  Also,  the EH theory serves as a foundational tool to study nonlinear phenomena in both astrophysics and cosmology.

The first EH black hole  appeared  as a magnetically charged black hole solution in~\cite{Yajima:2000kw} in the Einstein-Euler-Heisenberg theory twenty-five years ago.  Subsequent studies focused  on deriving electrically charged black hole solutions \cite{Ruffini:2013hia}, rotating black hole solutions \cite{Breton:2019arv,Amaro:2022yew}, and black hole solutions found from  modified gravity theories \cite{Guerrero:2020uhn,Nashed:2021ctg}.  Inspired by the low-energy limit of string theory and the Lovelock theory, an extension of Einstein-Maxwell-dilaton (EMd) theory has been proposed when coupling the  dilaton to nonlinear EH term to give the dilaton-Euler-Heisenberg (dEH) black hole solution~\cite{Bakopoulos:2024hah}. We wish to point out that  the difference between EMd and dEH theories is clearly shown by the presence of  a coupling term  $f(\phi)$ in Eq.\eqref{action}.
After that, some authors have  investigated the effects of the dEH black holes on particle motion and gravitational lensing phenomenon~\cite{Yasir:2025npe,Vachher:2024fxs}, and black hole shadow analysis \cite{Xu:2024gjs,Myung:2025zxu}. Subsequently, Jiang et al. \cite{Jiang:2024njc} have constructed  (geometrically thin and optically thick) accretion disks around the dEH black holes.

On the other hand, quasinormal modes (QNMs) indicate  essential characteristics of dissipative systems.  Particularly, black holes act as thermodynamic, open dissipative systems that consume matter and energy, increasing their event horizon area (entropy), while radiating energy through Hawking radiation or gravitational waves. Hence, black holes are characterized by (non-conservative) QNM oscillations and can exhibit dissipative effects like superradiance, where they lose angular momentum.
Also, we note that the QNMs appear from the ringdown phase of gravitational waves when binary black hole merge~\cite{Berti:2007}.  Unlike normal modes, it is worth noting that the eigenfunctions of QNMs do not form a complete set and are not normalizable~\cite{Nollert:1999}. Actually, the QNMs exhibit complex frequencies with real part representing vibration frequency and imaginary part indicating decay time scale.  Studying on the QNMs is crucial for inferring the mass and spin of black holes  as well as for  testing the no-hair theorem~\cite{Berti:2006,Berti:2007b,Isi:2019}. For  horizonless compact objects (neutron stars), the QNMs may reveal echoes in the ringdown phase, implying  an evidence for their existence ~\cite{Cardoso:2017,Cardoso:2016,Cardoso:2019}. Furthermore, the QNMs constrained modified gravity theories~\cite{Wang:2004,Blazquez-Salcedo:2016,Franciolini:2019,Aragon:2021,Liu:2021,Karakasis:2022,Cano:2022,Gonzalez:2022upu,Zhao:2022gxl} and they were used to clarify  the (in) stability of black holes  under metric and field perturbations~\cite{Jaramillo:2021,Cheung:2022,Ishibashi:2003,Chowdhury:2022zqg}. 
For the study of various black hole spacetimes  including those under the influence of magnetic fields, regular black holes, high-dimensional noncommutative black holes, and noncommutative black holes coupled to Einstein's tensor, many researchers have explored the impacts of different black hole  parameters on the  QNM behavior and stability~\cite{Yan:2020hga,Wu:2015fwa,Wu:2018xza,Yan:2020nvk}.

At this stage, we would like to mention that the separation of metric perturbations was done when choosing the Regge-Wheeler gauge. The tensor-dilaton perturbations are divided into axial perturbations, which gain a factor of $(-1)^{l+1}$ under parity inversion, and polar perturbations, gaining a factor of $(-1)^{l}$.  We stress that the polar perturbation includes the dilaton with $l\ge0$ and thus, we  have to  perform the stability analysis of the dEH black holes by computing QNM frequencies for  the metric tensor-dilaton perturbations. On the other hand, the axial perturbation includes the metric perturbation with $l\ge2$ only.  The stability analysis for the dilaton with $l\ge 0$  is important for checking stability of the dEH black hole with dilaton hair.  
It is noted    that the stability analysis for the dEH black holes is an important issue
since it determines their viability in representing realistic astrophysical configurations.
One expects to find  that the dEH black holes  with dilaton hair is stable against polar (metric-dilaton) perturbations with $l=0,1,2,3$.

This paper is organized as follows. In Sec.\ref{sec2}, we mention 
the considering  action and the dEH black hole solutions. In Sec. \ref{sec3}, we
describe  how to solve  four  coupled equations for the polar perturbations.
In Sec. \ref{sec4}, we describe  the direct integration method and  matrix values continued fraction method  to compute QNM frequencies. We present our numerical results for the polar (metric-dilaton) perturbations around the dEH black holes in Sec. \ref{sec5}. Finally,  we summarize  our results briefly in Sec. \ref{sec6}.
 
\section{Black hole Solutions}
\label{sec2}
\subsection{Action and Black Hole Solutions}
Considering  the low-energy limit of  string theory and Lovelock theory, Bakopoulos et al.~\cite{Bakopoulos:2024hah}  recently proposed the dilaton-Einstein-Maxwell (dEH) theory
(Einstein-Maxwell-dilaton (EMd) theory plus a dilaton coupling  to 
the nonlinear EH term) as 
\begin{eqnarray}
S=\frac{1}{16\pi}\int  d^4x\sqrt{-g}\Big[R-2\nabla^\mu\phi\nabla_\mu\phi-e^{-2\phi}F^2-f(\phi)\left(2\alpha F^\mu_{~\nu} F^\nu_{~\rho} F^\rho_{~\delta} F^\delta_{~\mu}-\beta F^4\right)\Big],\label{action}
\end{eqnarray}
where $R$ denotes the scalar curvature, $f(\phi)$ is a dilaton coupling function, Maxwell term  ($F^2=F_{\mu\nu}F^{\mu\nu}$), and $F^4=(F_{\mu\nu}F^{\mu\nu})^2$, where $F_{\mu\nu}$ stands for the field strength $F_{\mu\nu}=\partial_\mu A_{\nu}-\partial_{\nu}A_{\mu}$. Here $\alpha$ and  $\beta$ represent two dilaton coupling constants for this theory.  
First of all, let us describe famous known black hole solutions obtained from  the action \eqref{action}. It admitted the GMGHS black hole~\cite{Gibbons:1987ps,Garfinkle:1990qj} as an exact solution when $f(\phi)=0$. 
 For $\alpha=\beta=1/2$ and  $f(\phi)=2$, one found the dyonic Reissner-Nordstr\"om (dRN) black hole solution~\cite{Liu:2019rib}
\begin{eqnarray}
&&ds^2_{\rm dRN}=\bar{g}_{\mu\nu}dx^\mu dx^\nu=-f(r) dt^2+\frac{dr^2}{f(r)} +r^2d\Omega^2_2,  \nonumber \\
&& f(r)=1-\frac{2M}{r}+\frac{P^2}{r^2}+\frac{Q^2}{r^2} {}_2F_1\Big[\frac{1}{4},1,\frac{5}{4};-\frac{4P^2}{r^4}\Big], \label{dRN-bh}\\
&& \bar{A}=\bar{v}(r,P,Q)dt+P\cos\theta d\varphi \nonumber
\end{eqnarray}
with $ {}_2F_1[\cdots]$ the hypergeometric function.
For large $r$, the metric function takes the series form
 \begin{equation} \label{f-series}
  f(r)=1-\frac{2M}{r}+\frac{Q^2+P^2}{r^2}-\frac{4Q^2P^2}{5r^6}+\frac{16Q^2P^4}{9r^{10}}-\frac{64 Q^2P^6}{13 r^{14}}+\cdots,
 \end{equation}
 where the first-three terms represent the standard form for the dyonic black hole solution.
 We find the RN black hole for $Q=0$ in the dRN black hole solution. 

Varying the action \eqref{action} with respect to $g_{\mu\nu}$, $\phi$, and $A_\mu$, we can obtain the three  equations
\begin{eqnarray}
&&R_{\mu\nu}-\frac{1}{2}R g_{\mu\nu}=2{\partial _\mu}\phi{\partial_ \nu}\phi-g_{\mu\nu}{\partial }^\mu\phi {\partial }_\mu \phi+2T_{\mu\nu},\label{eqG}\\
&&\Box  \phi +\frac{1}{2}e^{-2\phi}F^2-\frac{df(\phi)}{d\phi}\left(\frac{\alpha}{2} F^\mu_{~\nu} F^\nu_{~\gamma} F^\gamma_{~\delta} F^\delta_{~\mu}-\frac{\beta}{4} F^4\right)=0,\label{eqKG}\\
&&{\partial_ \mu}\Big[\sqrt{-g}\left(e^{-2\phi}F^{\mu\nu}+f(\phi)(4\alpha F^\mu_{~\kappa} F^\kappa_{~\lambda} F^{\nu\lambda}-2\beta F^2 F^{\mu\nu})\right)\Big]=0,\label{eqMaxwell}
\end{eqnarray}
where $T_{\mu\nu}$ is the  energy-momentum  defined by
\begin{eqnarray}
T_{\mu\nu}&=&e^{-2\phi}(F^{\alpha}_\mu F_{\nu \alpha}-\frac{1}{4}g_{\mu\nu }F^2)\nonumber\\
&&+f(\phi)\left ( 4\alpha F^\alpha _\mu F^\beta_\nu F^\eta_\alpha F_{\beta\eta} -\frac{1}{2}\alpha g_{\mu\nu} F^\alpha_\beta F^\beta _\gamma F^\gamma _\delta F^\delta_\alpha -2\beta F^\xi_\mu F_{\nu\xi}F^2+\frac{1}{4}g_{\mu\nu}\beta F^4\right).\nonumber
\end{eqnarray}
We go further by introducing  the  dilaton coupling function $f(\phi)$ as
\begin{eqnarray}
 f(\phi)=-(3\text{cosh}(2\phi)+2)\to -\frac{3}{2}\left(e^{-2\phi}+e^{2\phi}+\frac{4}{3}\right).\label{phi}
\end{eqnarray}
Eqs.\eqref{eqG}-\eqref{eqMaxwell} admitted  the  magnetically charged black
hole solution \cite{Bakopoulos:2024hah}
\begin{eqnarray}
 ds^2&=&-H(r) \,dt^2 + \frac{1}{H(r)} \, dr^2 + R(r)^2 \,( d\theta^2 + \sin^2 \theta \, d\varphi^2),\label{oldmetric}\\
H(r)&=&1-\frac{2 M}{r}-\frac{2\epsilon Q_m^4}{r^3(r-Q_m^2/M)^3}, \quad
R(r)^2=r\left(r-\frac{Q_m^2}{M}\right),\nonumber\\
\bar{\phi} (r)&=&-\frac{1}{2}\ln \left(1-\frac{Q_m^2}{M r}\right), \quad A_\varphi= Q_m\cos\theta,\label{Qm}\label{fRsolution}
\end{eqnarray}
where $M$ and $Q_m$ are the mass and magnetic charge of this black hole with  $\epsilon=\alpha-\beta$.
Note  that  for $\epsilon=0(\alpha=\beta=0)$,  this solution \eqref{fRsolution} becomes  the  GMGHS black holes \cite{Gibbons:1987ps,Garfinkle:1990qj}.   Further, it  reduces to  the Schwarzschild solution for $Q_m=0$ but one could not find RN black hole from this solution. The related works in ~\cite{Ferrari:2000ep,Chen:2004zr,Shu:2004fj,Karimov:2018whx} have generated enormous interest in the GMGHS black holes. In addition,  Ref.~\cite{Bakopoulos:2024hah} showed that for $\epsilon=1$, the solution \eqref{fRsolution}  can describe a black hole with  single horizon, whereas for $\epsilon=-1$, the number of horizons  ranges from two to none. It  implies that the magnetically charged black hole possesses different horizon structures. Therefore, we  have  to consider both $\epsilon=1$ and $\epsilon=-1$.

To make a further progress,  it would be better to  transform  the metric \eqref{oldmetric} and dilaton solution \eqref{fRsolution} into a spherically symmetric dilaton-Euler-Heisenberg (dEH) black hole with dilaton hair as 
\begin{eqnarray}
  ds^2_{\rm dEH} &\equiv& \bar{g}_{\mu\nu}dx^\mu dx^\nu = -A(r)dt^2 + \frac{1}{B(r)} dr^2 + r^2 (d\theta^2 + \sin^2 \theta d\varphi^2),\label{metric}\\
A(r)&=&1-\frac{4 M^2}{Q_m^2+\sqrt{Q_m^4+4 M^2 r^2}}-\frac{2 \epsilon Q_m^4}{r^6},\nonumber\\
B(r)&=&1
- \frac{Q_{m}^{4} + 4M^{2}r^{2}}{r^2(Q_{m}^{2} + \sqrt{Q_{m}^{4} + 4M^{2}r^{2}})} +\frac{Q_m^4}{4M^2r^2}
- \frac{\epsilon Q_{m}^{4}(Q_{m}^{4} + 4M^{2}r^{2})}{2M^2r^{8}},\nonumber\\
\bar{\phi} (r)&=&-\frac{1}{2}\ln \left(\frac{\sqrt{Q_m^4+4 M^2 r^2}-Q_m^2}{\sqrt{Q_m^4+4 M^2 r^2}+Q_m^2}\right)\label{solution}.
\end{eqnarray}

Eq.\eqref{metric} is obtained from Eq.\eqref{oldmetric} via a coordinate transformation, performed following Ref.~\cite{Bakopoulos:2024hah}, which maps the unphysical coordinates to physical ones. We will use the metric given by Eq.\eqref{metric} to compute the QNM frequencies. For $\epsilon=-1$ and $M=1$ the dEH black hole becomes extremal near $Q_m\approx0.826$, for which the QNM frequencies are computed in the regime $Q_m\textless0.826$.

\subsection{Ghost analysis of the vector perturbations}

To ensure the physical viability of the dEH black hole, we examine the kinetic sector of the vector perturbations. A ghost-free theory requires the effective kinetic coupling $K(\mathcal{F}) = \partial \mathcal{L}_{vec} / \partial \mathcal{F}$ to remain negative throughout the exterior spacetime, where $\mathcal{F} \equiv F_{\mu\nu}F^{\mu\nu}$.
Here the vector Lagrangian density from the general action \eqref{action} is 
\begin{equation}
    \mathcal{L}_{vec} = -e^{-2\phi} \mathcal{F} - f(\phi) \left( 2\alpha F^{\mu}_{~\nu}F^{\nu}_{~\rho}F^{\rho}_{~\delta}F^{\delta}_{~\mu} - \beta \mathcal{F}^2 \right).
    \label{L_vec_original}
\end{equation}

For the magnetically charged black hole background, the only non-zero component of the gauge field is $A_\varphi = Q_m \cos\theta$, which yields the non-vanishing field strength components $F_{\theta\varphi} = -F_{\varphi\theta} = Q_m \sin\theta$. In this case, the mixed tensor $F^{\mu}_{~\nu} = g^{\mu\lambda}F_{\lambda\nu}$ can be represented as a $4\times4$ matrix with non-zero elements only in the $\theta$-$\varphi$ subsector:
\begin{equation}
    F^{\mu}_{~\nu} = \begin{pmatrix}
    0 & 0 & 0 & 0 \\
    0 & 0 & 0 & 0 \\
    0 & 0 & 0 & F^{\theta}_{~\varphi} \\
    0 & 0 & F^{\varphi}_{~\theta} & 0
    \end{pmatrix}.
\end{equation}
By direct matrix multiplication, the square of this matrix is:
\begin{equation}
    (F^2)^{\mu}_{~\nu} = \begin{pmatrix}
    0 & 0 & 0 & 0 \\
    0 & 0 & 0 & 0 \\
    0 & 0 & F^{\theta}_{~\varphi}F^{\varphi}_{~\theta} & 0 \\
    0 & 0 & 0 & F^{\varphi}_{~\theta}F^{\theta}_{~\varphi}
    \end{pmatrix},
\end{equation}
which leads to the standard field invariant $\mathcal{F} = F_{\mu\nu}F^{\mu\nu} = -\text{Tr}(F^2) = -2 F^{\theta}_{~\varphi}F^{\varphi}_{~\theta}$. 
Similarly, for the four-product term $F^{\mu}_{~\nu}F^{\nu}_{~\rho}F^{\rho}_{~\delta}F^{\delta}_{~\mu} = \text{Tr}(F^4)$, we have:
\begin{equation}
    \text{Tr}(F^4) = 2 \left(F^{\theta}_{~\varphi}F^{\varphi}_{~\theta}\right)^2 = 2 \left( -\frac{1}{2}\mathcal{F} \right)^2 = \frac{1}{2} \mathcal{F}^2.
    \label{trace_identity}
\end{equation}

Substituting the identity \eqref{trace_identity} and $F^4 = \mathcal{F}^2$ back into the original vector Lagrangian \eqref{L_vec_original}, we obtain:
\begin{align}
    \mathcal{L}_{vec} &= -e^{-2\phi} \mathcal{F} - f(\phi) \left[ 2\alpha \left(\frac{1}{2}\mathcal{F}^2\right) - \beta \mathcal{F}^2 \right] \nonumber \\
    &= -e^{-2\phi} \mathcal{F} - (\alpha - \beta) f(\phi) \mathcal{F}^2 \nonumber \\
    &= -e^{-2\phi} \mathcal{F} - \epsilon f(\phi) \mathcal{F}^2,
    \label{L_vec_final}
\end{align}
where $\epsilon = \alpha - \beta$. This completes the derivation of the simplified vector Lagrangian.

The kinetic coefficient is then obtained by differentiating Eq. \eqref{L_vec_final} with respect to $\mathcal{F}$:
\begin{equation}
    K(r) = \frac{\partial \mathcal{L}_{vec}}{\partial \mathcal{F}} = -e^{-2\phi(r)} - 2\epsilon f(\phi(r)) \bar{\mathcal{F}},
\end{equation}
with the background field strength $\bar{\mathcal{F}} = 2Q_m^2/r^4$. Substituting this into the expression for $K(r)$, we obtain:
\begin{equation}
    K(r) = -e^{-2\phi(r)} - \frac{4\epsilon f(\phi) Q_m^2}{r^4}.
\end{equation}

The absence of ghosts requires that $K(r)$  maintain a consistent sign (negative, in our convention) throughout the exterior spacetime. Based on the dilaton coupling function $f(\phi)$ defined in Eq. \eqref{phi} (which is strictly negative, $f(\phi) < 0$), we discuss the following two cases:
\begin{itemize}
    \item \textbf{Case $\epsilon = -1$}: The term $-4\epsilon f(\phi) = 4f(\phi)$ is strictly negative. $K(r)$ is thus a sum of two negative terms, ensuring the theory is unconditionally ghost-free.
    \item \textbf{Case $\epsilon = 1$}: The second term is positive, which could potentially lead to $K > 0$ near the horizon where the magnetic field is strongest. Evaluating the condition at the horizon reveals that the exterior spacetime remains ghost-free for $Q_m < 0.64$. 
\end{itemize}

Since the numerical results in this work are obtained within the parameter range $Q_m < 0.64$ for $\epsilon = 1$, the investigated dEH black hole models are physically consistent and free from ghost instabilities.



\section{Polar perturbations}
\label{sec3}

In this section, we wish to derive the linearized equations governing the polar (even-parity)  perturbations around  the dEH black hole given by  Eq.(\ref{metric}). Due to the presence of the dilatonic coupling to the nonlinear  EH  term, one expects that  the polar equations  take   a system of coupled equations for the metric and dilaton perturbations.

Now, we are in a position  consider the perturbations around the dEH black hole background
\begin{eqnarray}
g_{\mu\nu}=\bar{g}_{\mu\nu}+h_{\mu\nu},\quad \phi=\bar{\phi}+\delta\phi,
\end{eqnarray}
where  $h_{\mu\nu}$ represents the perturbed metric with $h_{\mu\nu}\ll\bar{g}_{\mu\nu}$ and $\delta \phi$ denotes a perturbed dilaton, being functions of $(t,r,\theta,\varphi)$. 
The linearized  equations can be further simplified by taking Fourier transformation on these quantities and  expanding them in terms of tensor and scalar spherical harmonics.
For a scalar perturbation, it is given by 
\begin{eqnarray}
\delta\phi = \int  d{\omega}\, e^{-i\omega t}\, \sum_{l,m}\,\delta\phi_1(r)\, Y_{l}^{m}(\theta,\varphi).\label{scalarpert}
\end{eqnarray}
Imposing the Regge–Wheeler gauge~\cite{Regge:1957}, we expand the metric perturbations in terms of tensor spherical harmonics for the polar perturbation  with Zerilli variables (four radial modes of $H_0, H_1, H_2, K$)
\begin{eqnarray}
  h_{\mu\nu} =  \int  d{\omega}e^{-i\omega t}  \sum_{l,m} 
\begin{bmatrix}
A(r)H_0(r) & H_1(r) & 0 & 0 \\
H_1(r) & \frac {H_2(r)}{B(r)} & 0 & 0 \\
0 & 0 & r^2 K(r) & 0 \\
0 & 0 & 0 & r^2\sin^2 \theta  K(r)
\end{bmatrix}
Y_{l}^{m}(\theta,\varphi)\label{eqeven}.
\end{eqnarray}
Hereafter,  we  choose $m=0$  for simplicity, as all perturbation equations are independent of $m$ \cite{Regge:1957}.

The linearized theory is usually described  by axial and polar  equations  which can be obtained by expanding  Eqs.\eqref{eqG} and \eqref{eqKG} up to the first order in  $h_{\mu\nu}$ and $\delta\phi$~\cite{Myung:2018jvi}. 
The polar part  is composed of four coupled equations for Zerilli variables  and dilaton  with angular momentum quantum number $l\ge0$ as
\begin{eqnarray}
K'(r) &=& 
 \left(\frac{A'}{2 A}-\frac{1}{r}\right)K(r)+\frac{i  l (l+1)}{2 r^2 \omega }H_1(r)+\frac{H_0(r)}{r}-\frac{2  \bar{\phi} '}{r}\delta\phi _1(r),\label{eq:Kp}\\
H_1'(r) &=& 
- \frac{i\omega\, H_0(r)}{B} 
- \frac{i\omega\, K(r)}{B}-  \left( \frac{A'}{2A} + \frac{B'}{2B} \right) H_1(r) ,\label{eq:H1p}\\
H_0'(r) &=& 
 \left(\frac{1}{r} -\frac{A'}{A}\right) H_0(r)+ \left(\frac{A'}{2 A}-\frac{1}{r}\right)K(r)+ \left(\frac{i l (l+1)}{2 r^2 \omega }-\frac{i \omega }{A}\right)H_1(r)+\frac{2 \bar{\phi} '}{r}\delta\phi _1(r),\label{eq:H0p}\\
\delta\phi_1''(r)  &=& \left(\frac{5 A'}{2 A r}+\frac{2 A' \phi '}{A}-\frac{\omega ^2}{A B}-\frac{3 B'}{2 B r}+\frac{l(l+1)}{B r^2}-\frac{12 \epsilon  e^{-2 \bar{\phi} } Q_{m}^4}{B r^8}+\frac{4 e^{-2 \bar{\phi} } Q_{m}^2}{B r^4}\right) \delta\phi _1(r)\nonumber\\
&&+\left(-\frac{A'}{2 A}-\frac{B'}{2 B}\right)\delta\phi _1'(r) + \left(-\frac{4 r A' \bar{\phi} '}{A}-\frac{2 e^{-2 \bar{\phi} } Q_{m}^2}{B r^3}\right)K(r). \label{eq:phi1pp}
\end{eqnarray}
From  ($\theta,\phi$)-component of the perturbation equation, we have \( H_2(r) = H_0(r) \) and from the $(r,r)$-component, we obtain an  algebraic identity
\begin{eqnarray}
&&\Big[2 r^2 \left(2 \omega ^2-B A''\right)+4 A^2+\frac{B r^2 (A')^2}{A}-2 A \left(r A'+l(l+1)\right)\Big]K(r)- \Big(4 r \omega ^2-l(l+1) A'\Big)\frac{iB H_1(r)}{\omega} \nonumber\\
&&-\Big(4 A^2-2 B r A'-2 A l(l+1)\Big)H_0(r)+4  B  \bar{\phi} ' \left(r A'-2 A\right) \delta\phi_{1}(r)-8 A B r \bar{\phi}' \delta\phi_{1}'(r)=0.
\label{eq:constraint}
\end{eqnarray}
The polar equations reduce to two coupled second order equations and  the relevant derivation will be found  in Appendix \ref{appendix-A}.
The Zerilli equation is given by
\begin{eqnarray}
&&\Big[\frac{d^2}{dr_{*}^{2}}+\omega^2\Big]\hat{K}(r)=V_{KK}\hat{K}(r)+V_{KS}\hat{S}(r),
\label{eq:Zerilli}
\end{eqnarray}
\begin{alignat}{2}
V_{KK}=
&\Big\{-12&& A^6+4 A^5 \Big[r A'+3 B+4 l (l+1)\Big]-B^3 r^4 A'^2 A''\nonumber\\
&-A^4 \Big[&&l (l+1) \left(4 r A'+7 l (l+1)\right)+2 B \left(-5 r^2 A''+r A'+6 l (l+1)\right)\Big]\nonumber\\
&+A^3 \Big[&&l^2 (l+1)^2 \left(r A'+l^2+l\right)+B \left(l (l+1) \left(3 l (l+1)-7 r^2 A''\right)+l (l+1) r A'-4 r^2 \left(A'\right)^2\right)\nonumber\\
& &&+2 B^2 r \left(r \left(A''' r-5 A''\right)-A'\right)\Big]\nonumber\\
&+A B^2&& r^2 \Big[2 B r^2 \left(A''\right)^2-l (l+1) \left(A'\right)^2+r \left(A'\right)^3+B r A' \left(A''-A''' r\right)\Big]\nonumber\\
&+A^2 B&& r \Big[l (l+1) r \left(l (l+1) A''+2 \left(A'\right)^2\right)\nonumber\\
& &&+B \left(4 r \left(A'\right)^2+l (l+1) r \left(5 A''-A''' r\right)+A' \left(-3 r^2 A''+l(l+1)\right)\right)\Big]\Big\}\nonumber\\
&\times&&\frac{1}{r^2 \left(-2 A^2+B r A'+A l (l+1)\right)^2},
\label{VKK}
\end{alignat}
\begin{alignat}{1}
V_{KS}=&\frac{8 (A-B) \sqrt{A B} \left(4 A^3+B r^2 \left(A'\right)^2+A r \left(A' \left(-B+l^2+l\right)-B r A''\right)-2 A^2 \left(r A'+l^2+l\right)\right)}{r Q_m^2 \left(B r A'+A \left(-2 A+l^2+l\right)\right)^2}.
\label{VKS}
\end{alignat}
On the other hand, the dilaton  equation takes the form
\begin{eqnarray}
&&\Big[\frac{d^2}{dr_{*}^{2}}+\omega^2\Big]\hat{S}(r)=V_{SS}\hat{S}(r)+V_{SK}\hat{K}(r)+U_{SK}\frac{d\hat{K}(r)}{dr_{*}},
\label{eq:Scalar}
\end{eqnarray}
where 
\begin{alignat}{1}
V_{SS}=&\frac{8 A (A-B) \left(2 B r^4 A' \bar{\phi} '+A e^{-2 \bar{\phi} } Q_{m}^2\right)}{r^5 \bar{\phi} ' \left(B r A'+A \left(-2 A+l^2+l\right)\right)}+\frac{5 B A'-3 A B'}{2 r}+2 B A' \bar{\phi} '+\frac{A l (l+1)}{r^2}\nonumber\\
&-\frac{12 A \epsilon  e^{-2 \bar{\phi} } Q_{m}^4}{r^8}+\frac{4 A e^{-2 \bar{\phi} } Q_{m}^2}{r^4},
\label{VSS}
\end{alignat}
\begin{alignat}{1}
&V_{SK}=\nonumber\\
&\frac{A e^{-2 \bar{\phi} } Q_{m}^2 \left(2 B^2 r^2 A''+\left(-2 A+l^2+l\right) \left(B r A'+A \left(-2 A+2 B+l^2+l\right)\right)\right) \left(2 B r^4 e^{2 \bar{\phi} } A' \bar{\phi} '+A Q_{m}^2\right)}{r^6 \sqrt{A B} \bar{\phi} ' \Big(B r A'+A \left(-2 A+l^2+l\right)\Big)},
\label{VSK}
\end{alignat}
\begin{alignat}{1}
U_{SK}=&\frac{4 B A' Q_{m}^2}{r}+\frac{2 A e^{-2 \bar{\phi} } Q_{m}^4}{r^5 \bar{\phi} '}
\label{USK}
\end{alignat}

\section{Computation Scheme for Quasinormal Modes}
\label{sec4}

\subsection{Two Computation Methods}
\label{subsubsec4-A-1}

In the case of polar perturbations around  the dEH black hole, the linearized theory consists of four  coupled first-order and second-order equations involving both metric and dilaton degrees of freedom.  Because the quasinormal boundary conditions couple strongly across the different (axial, polar) perturbations, a standard single-field technique (WKB approximation) is  not suitable for handling  our case. Therefore, we have to introduce  two robust numerical approaches specifically suited for coupled systems: direct integration   and continued fraction methods. These  methods provide independent determinations of the QNM  frequencies and allow us to verify the stability of dEH black holes and accuracy of our computation.
We wish to mention briefly   two methods.

\subsubsection{Matrix-valued direct Integration Method}
\label{subsubsec4-A-1-new}

To implement the matrix-valued direct integration method~\cite{Pani:2013}, the coupled system (\ref{eq:Kp})-(\ref{eq:phi1pp}) is reformulated into a set of four first-order equations. In the polar perturbations, two of  three metric  perturbations  are considered as independent perturbations.   For instance, a variable $H_2(r)$ is not an independent one and thus, it is determined via a constraint equation by  other variables of   $H_1(r)=\omega R_1(r)$ and $K(r)$.  The dilaton perturbation is governed by a second-order equation\eqref{eq:phi1pp}, which can be recast into a system of two first-order equations.
Thus, to prevent the system from being over-constrained and to obtain a uniquely determined solution, one needs to find  the first-order system as
\begin{eqnarray}
\frac{d}{dr}\Psi_j+V_j\Psi_j=0,
\end{eqnarray}
where $\Psi_j$ and $V_j$ are column vectors with $\Psi_j\equiv(R_{1},K,\delta\phi_1,\delta\phi_1')$ and $V_j$ given by Eqs.(\ref{eq:Kp})-(\ref{eq:phi1pp}). Specifically, we note that $V_3 =(0,0,0,-1)$. For a concrete example, see Ref.\cite{Blazquez-Salcedo:2016enn}, as well as recent related studies in Refs.\cite{Gu:2025lyz}.

In this case, we  may solve the coupled system of ordinary differential equations by enforcing the appropriate boundary conditions at the event horizon ($r \to r_h$) and spatial infinity ($r \to \infty$).
In the near-horizon, the physical solution must be purely ingoing, while  it must be purely outgoing at  infinity. The asymptotic form  of the perturbation vector $\Psi_j$ is given by
\begin{eqnarray}
\Psi_j\propto\begin{cases}e^{-i\omega r_*},\quad r\sim r_h,\\
e^{i\omega r_*},\quad r\to\infty,&\end{cases}
\end{eqnarray}
where the tortoise coordinate $r_*$  is defined by
\begin{eqnarray}
\frac{dr_*}{dr}=\frac{1}{\sqrt{A(r)B(r)}}.
\end{eqnarray}
Here, $A(r)$ and $B(r)$ denote the metric functions appearing in Eq.\eqref{metric}. Substituting the asymptotic expansions of $\Psi_{j}$ at the horizon and at infinity into the coupled equations, we obtain four recurrence relations
\begin{eqnarray}
&&\Psi_{j}(r\to r_{h})\sim e^{-i\omega r_{*}}\sum_{k=0}^{N}\psi_{j,k}(r-r_{h})^{k+p_j},\\
&&\Psi_{j}(r\to \infty)\sim e^{i \omega r_{*}}\sum_{k=0}^{N}\psi_{j,k}(\frac{1}{r})^{k+p_j}.
\end{eqnarray}
Here, $\Psi_j$ represents the set $\{R_{1},K,\delta\phi_1,\delta\phi_1'\}$, and the associated exponents are given by $p_j=\{-1,0,0,0\}$. Through the recurrence relations, all expansion coefficients at the horizon and at infinity can be expressed in terms of two coefficients, respectively. We may integrate the coupled equations outward from the horizon and inward from infinity to a matching point.  The QNM frequencies ($\omega$) could be  obtained by imposing the consistency of  two solutions at the matching  point as  
\begin{eqnarray}
\det(X)|_{r=r_\mathrm{m}}=0.
\end{eqnarray}
Here, $X$ is  $4\times 4$ matrix constructed  by performing 4 integrations of coupled equation for $\Psi_j$ from the horizon to infinity. 
The matrix $X$ is a function of  the characteristic frequencies $\omega = \omega_R + i \omega_I$  and  whose determinant $\det(X)$ is used to extract the QNM frequencies. For the matching radius, we choose $r_m\sim 4r_h$.
The QNM frequencies  are identified as the roots which  satisfy two boundary conditions simultaneously, and they are typically found by minimizing the determinant of the matching matrix.

\subsubsection{Matrix Values Continued Fraction Method}
\label{subsubsec4-A-2}

To verify the results obtained via the previous direct integration, we adopt  the matrix values continued fraction method (MVCFM), so-called a generalized Leaver’s method~\cite{Leaver:1985ax,Leaver:1990zz,Berti:2009kk} for coupled systems~\cite{Rosa:2011my,Pani:2013pma,Pani:2012bp,Nomura:2021efi,Antoniou:2024jku}. This method is particularly effective for slowly converging potentials. We start by factoring out the asymptotic behaviors at the horizon and infinity. We follow an ansatz~\cite{Pani:2013} that satisfies the required boundary conditions at both the horizon and infinity:
\begin{eqnarray}
Y_i = e^{-i \omega r_*} \, r^{-\nu} \, e^{q r} \sum_n a_n^{(i)} F(r)^n,
\end{eqnarray}
Where $q=i\omega$ and $v=-2i\omega k_H$, and $k_H=\frac{1}{\sqrt{A'(r_+)B'(r_+)}}$. $F(r)$ is chosen such that $F(r_+)=0$ and $F(\infty)=1$, and is given by $F(r)=1-\frac{r_+}{r}$.
Substituting this series into the coupled equations leads to a matrix three-term recurrence relation for the expansion coefficients $a_n$:
\begin{eqnarray}
\alpha_0 a_1 + \beta_0 a_0 = 0, \qquad n = 0,
\label{eq:recu0}
\end{eqnarray}
\begin{eqnarray}
\alpha_n a_{n+1} + \beta_n a_n + \gamma_n a_{n-1} = 0, \qquad n > 0,
\label{eq:recu1}
\end{eqnarray}
where $a_n$ is a column vector containing the $n$-th coefficients of coupled fields, and $\alpha_n, \beta_n, \gamma_n$ form  $4\times4$ matrix, depending on the QNM frequency $\omega$ and  black hole parameters.
To address this issue clearly, we choose  a procedure for the continued fraction method as suggested in~\cite{Pani:2013}.
For our purpose, we may introduce  the ladder matrix $\mathbf{R}_n^+$ 
\begin{eqnarray}
 \mathbf{a}_{n+1} = \mathbf{R}_n^+ \mathbf{a}_n.
\end{eqnarray}
Making use of  the ladder matrix, Eq.\eqref{eq:recu0} can be reformulated as $\mathbf{M} \mathbf{a}_{0}=0$, where the matrix $\mathbf{M}$ is given by
\begin{eqnarray}
 \mathbf{M} \equiv \boldsymbol{\beta}_{0} + \boldsymbol{\alpha}_{0}\mathbf{R}_0^+.
\end{eqnarray}
Taking $n\rightarrow n+1$ in Eq.~\eqref{eq:recu1} and substituting the ladder matrix into the recurrence relation yields
\begin{eqnarray}
 \mathbf{R}_n^+ = - \left[ \boldsymbol{\beta}_{n+1} + \boldsymbol{\alpha}_{n+1} \mathbf{R}_{n+1}^+ \right]^{-1} \boldsymbol{\gamma}_{n+1}.
 \label{eq:recun}
\end{eqnarray}
We note that the ladder matrix $ \mathbf{R}_0^+ = - \left[ \boldsymbol{\beta}_{1} + \boldsymbol{\alpha}_{1} \mathbf{R}_{1}^+ \right]^{-1} \boldsymbol{\gamma}_{1}$ is derived from Eq.~\eqref{eq:recu1} with $n=1$ (equivalently, from Eq.~\eqref{eq:recun}). The matrix $\mathbf{R}_1^+$ can then be obtained  via the recurrence relation, starting from the truncated matrix $\mathbf{R}_N^+$ at $n=N$ with a sufficiently large $N$.
The eigen-frequencies of QNMs are determined when finding the roots of $\omega$, satisfying 
\begin{eqnarray}
\mathbf{det M}=0.
\label{eq:detM}
\end{eqnarray} 
In most cases,  three-term recurrence relation is not easily  obtained.  Instead, we find  a recurrence relation with up to $N+1$ terms
\begin{eqnarray}
 \label{eq:recursion-1}
\left[\begin{array}{cccccc}
A_{1,0} & A_{1,1} & & & & \\
A_{2,0} & A_{2,1} & A_{2,2} & & & \\
A_{3,0} & A_{3,1} & A_{3,2} & A_{3,3} & & \\
\vdots & \ddots & \ddots & \ddots & & \\
A_{N-1,0} & \cdots &  A_{N-1,N-3} & A_{N-1,N-2} & A_{N-1,N-1} & \\
A_{N,0} & \cdots & A_{N,N-3} & A_{N,N-2} & A_{N,N-1} & A_{N,N}
\end{array}\right]
\begin{bmatrix}
a_0 \\
a_1 \\
a_2 \\
\vdots \\
a_{N-1} \\
a_N
\end{bmatrix}
= 0.
\end{eqnarray}
It is urgent to derive its three-term recurrence relation (enabling computation via the continued fraction method). Here,   Gaussian elimination will be used  to eliminate redundant terms, yielding the following relation: 
\begin{eqnarray}
 \label{eq:recursion-2}
\left[\begin{array}{cccccc}
\tilde{A}_{1,0} & \tilde{A}_{1,1} & & & & \\
\tilde{A}_{2,0} & \tilde{A}_{2,1} & \tilde{A}_{2,2} & & & \\
 & \tilde{A}_{3,1} & \tilde{A}_{3,2} & \tilde{A}_{3,3} & & \\
 & \ddots & \ddots & \ddots & & \\
 &  &  \tilde{A}_{N-1,N-3} & \tilde{A}_{N-1,N-2} & \tilde{A}_{N-1,N-1} & \\
 &  &   & \tilde{A}_{N,N-2} & \tilde{A}_{N,N-1} & \tilde{A}_{N,N}
\end{array}\right]
\begin{bmatrix}
a_0 \\
a_1 \\
a_2 \\
\vdots \\
a_{N-1} \\
a_N
\end{bmatrix}
= 0.
\end{eqnarray}
Obtaining all three-term recurrence relations is allowed  by  applying Gaussian elimination repeatably.  A relation comprising $N+1$ terms, for example, requires $N-2$ elimination iterations. In our work, the use of a 30th-order continued fraction method produces the recurrences with a maximum of 31 terms. The reduction to a three-term recurrence  for the highest-order case is therefore achieved through 28 Gaussian elimination steps.  Through the Gaussian elimination steps, we carry out  the procedure outlined in Eqs. \eqref{eq:recu0} to \eqref{eq:detM} to find  the roots of the eigen-frequency $\omega$ of QNMs. Convergence of the continued fraction method is monitored by the behavior of the successive differences of the QNM frequencies as the truncation order increases. Additional details on this convergence analysis are presented in Appendix \ref{appendix-CFM}. Finally, we would like to mention that this method is highly efficient for higher overtones and coupled systems, offering better precision and convergence than direct integration for a wide range of parameters.

\section{Results and Analysis}
\label{sec5}

Before we proceed, we mention the stability analysis of the infinite branches ($n=0,1,2,\cdots$) of scalarized  dRN black hole (\ref{dRN-bh}) in the Einstein-Maxwell-scalar theory with a quasi-topological term ($\alpha=\beta=1$) by  coupling a scalar function $f(\phi)=e^{\alpha \phi^2}$  to both Maxwell and quasi-topological terms~\cite{Myung:2020ctt}. These scalarized dRN black holes correspond to the dEH black hole with dilaton hair (\ref{metric}). 
It turned out that  the $n = 0$ branch of  scalarized dRN black holes are stable against the radial perturbation with $l=0$,  while other branches ($ n = 1,2$) of scalarized dRN  black holes are unstable. This suggests that the  dilaton propagating   around the dEH black holes is stable against the $l=0$ polar perturbation.

The stability analysis of the dEH black holes  will be performed by obtaining  QNM  frequency of $\omega=\omega_R+\omega_I$  for gravitational mode $\Psi_g \in(R_1,K)$ and dilaton mode $\Psi_d\in \delta \phi_1$ when solving the linearized coupled equations
with appropriate boundary conditions. These conditions are found  at the outer horizon: ingoing waves and at infinity: purely outgoing waves. For the black hole background (\ref{oldmetric}),
the radial perturbation for a minimally coupled massive scalar with $l=1$ was performed  in~Ref.\cite{Bakopoulos:2024hah}. 

Here, we briefly  describe the dilaton stability of the dEH black holes.

\subsection{$l=0$}

In this case, since two Zerilli variables  $R_1(r)$ and $K(r)$ are redundant, the dilaton is a physically propagating mode around the dEH black holes.
For $l=0$ (s-mode), the polar linearized equation is given entirely by the dilaton equation
\begin{eqnarray}
   \frac{d^{2}\Psi_d(r_{*})}{{dr_{*}}^{2}}+[\omega^{2}-V_{{\rm eff},l=0}(r)]\Psi_d(r_{*})=0,
   \label{wavefunc-1}
\end{eqnarray}
where its  potential takes the form
\begin{eqnarray}
V_{{\rm eff}, l=0}(r) &&=\frac{B A' \left(3 r^2 \phi '^2+1\right)}{2 r}-\frac{A e^{-2 \phi }}{2 r^8} \Big\{-r^7 e^{2 \phi} \Big[B' \left(1-r^2 \phi '^2\right)+2 r B \left(r^2 \phi '^2-2\right) \phi '^2\Big] \nonumber \\
&&+2 r^4 Q_{m}^2 \left(r \phi '-2\right)+6 \epsilon  Q_{m}^4 \Big[r \left(e^{4 \phi }-1\right) \phi '+2 \left(e^{4 \phi }+1\right)\Big]\Big\}.\label{Veffl=0}
\end{eqnarray}
Although Eq.\eqref{Veffl=0} is different from Eq.(6.5) of Ref.\cite{Bakopoulos:2024hah}, it leads to  Eq.(6.5) when the background solution is applied. However, the authors of Ref.\cite{Bakopoulos:2024hah} did not compute QNM frequencies based on Eq.\eqref{Veffl=0}.

\begin{figure}[H]
\centering
\subfigure[~$\epsilon=1$]{
\label{figVeffl0zeta1} 
\includegraphics[height=1.5in]{
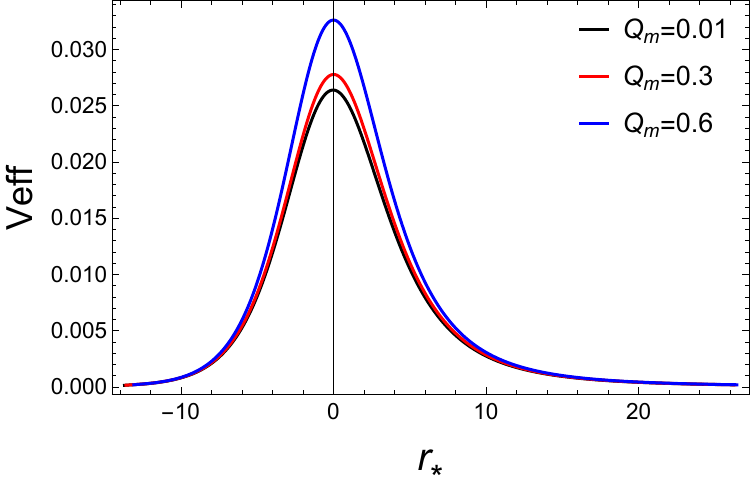}}
\quad
\subfigure[~$\epsilon=-1$]{
\label{figVeffl0zetam1} 
\includegraphics[height=1.5in]{
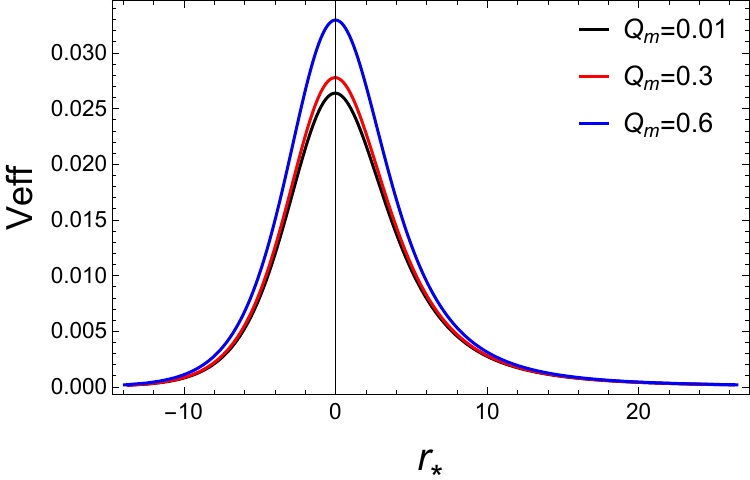}}
\caption{
Graphs of the potential for  the $l=0$ dilaton as function of $r_*$. }
\label{fig:l=0Vall}
\end{figure}

Fig.\ref{fig:l=0Vall} illustrates the behavior of the effective potential $V_{eff}(r_*)$ for the $l=0$ mode. As the magnetic charge $Q_m$ increases from 0.01 (black curve) to 0.6 (blue curve)~\cite{Bakopoulos:2024hah}, the width of the potential barrier expands, and the height of the barrier peak increases accordingly.
This evolution of the potential's shape will affect the characteristic frequency ($\omega_R$) and energy dissipation rate ($\omega_I$) of the quasinormal mode.
Since the effective potentials are positive definite, we expect to have the stability of this black hole under the $l=0$-mode  dilaton perturbation.

As shown in~\cite{Zhang:2025xqt}, there are QNM frequencies (AIM) for the  $l=0$ scalar mode propagating around the dEH black holes as 
\begin{eqnarray}
&&M\omega \approx 0.112132-0.105410i(\epsilon=1), \quad \omega \approx 0.112150-0.105337i(\epsilon=-1) \label{omegal03}
\end{eqnarray}
with  $M=1$ and $Q_m=0.3$. For  $M=1$ and $Q_m=0.6$, these are given by 
\begin{eqnarray}
&&M\omega \approx 0.117741-0.107542i(\epsilon=1), \quad M\omega \approx 0.118292-0.105480i(\epsilon=-1).  \label{omegal06}
\end{eqnarray}
We note that  these are slightly different from the $l=0$  dilaton  (AIM) in Table I.

\begin{table*}[h!]
\caption{Fundamental ($n=0$) QNM frequencies for the $l=0$-mode dilaton $\Psi_d$  and coupling parameters $\epsilon=\pm1$ around the  dEH black holes with $M=1$,  It indicates  values from  direct integration method (DI) and  asymptotic iteration method (AIM) across different $Q_m$ and the discrepancy between them.}
\label{table-A-1}
\resizebox{\linewidth}{!}{
\begin{tabular}{|c|c|c|c|c|c|c|c|} \hline
 & & \multicolumn{3}{|c|}{$\epsilon=1$}  &  \multicolumn{3}{|c|}{$\epsilon=-1$} \\ \hline
$l$ &  	$Q_m$ & DI  &  AIM  & $\Delta_{DA}$   &  DI  &  AIM   & $\Delta_{DA}$ \\ \hline
 \multirow{3}{*}{0}&  0.01  &   $0.112802-0.103022 i$  &  $0.110595-0.104480 i$   &1.73524\%   & $0.112802-0.103022 i$        & $0.110599-0.104489 i$  & 1.73621\%  \\ 
& 0.3 &  $0.117415-0.103837 i$  & $0.115845-0.105689 i$    &1.54887\%   & $0.117543-0.103772 i$   &  $0.115973-0.105614 i$  &   $1.54308\%$   \\ 
&  0.6 &  $0.131367-0.107965 i$ &$0.131794-0.110131 i$   &  $1.29207\%$ &$0.134008-0.106057 i$  & $0.135674-0.107850 i$ & $1.29207\%$ \\ \hline
\end{tabular}}
\end{table*}

\begin{figure}[H]
\centering
\subfigure[~$\omega_{R} $ vs $ Q_m$]{
\label{l=0figzeta1R} 
\includegraphics[height=1.5in]{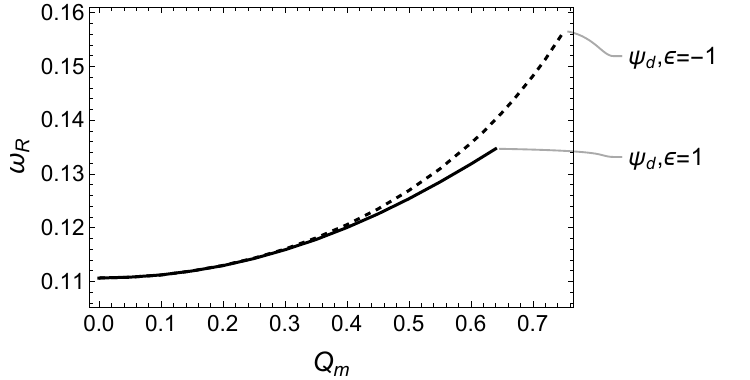}}
\quad
\subfigure[~$\omega_{I}$ vs $ Q_m$]{
\label{l=0figzeta1I} 
\includegraphics[height=1.5in]{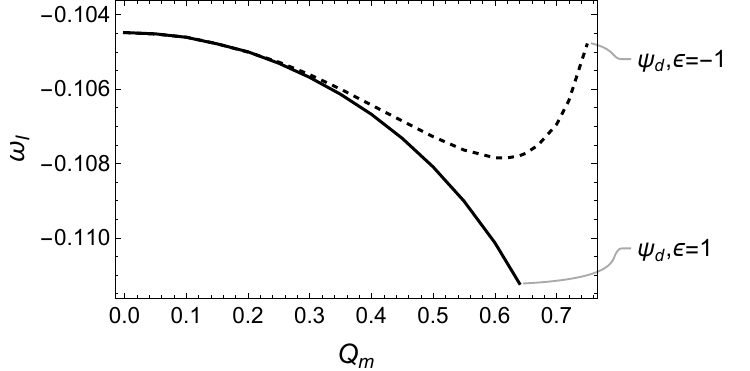}}
\caption{Variation of fundamental  QNM frequencies (real/imaginary parts) with $Q_m$ for the $l=0$ dilaton $\Psi_d$ around  dEH black holes with $M=1$. Solid lines are  for $\epsilon=1$, while dashed lines denote $\epsilon=-1$ case.}
\label{fig:l=0all}
\end{figure}

Table~\ref{table-A-1} and Fig.~\ref{fig:l=0all} together reveal the behavior of fundamental QNM frequencies ($n=0$) with respect to magnetic charge $Q_m$ and parameter $\epsilon$. Clearly, the frequencies $\omega_R$ and $\omega_I$ obtained by two methods (direct integration and asymptotic iteration methods ) are in very close agreement, with relative errors $\Delta _{DA}$ all below $2\%$. It supports the precision and reliability of these QNM frequences used for analysis.

In Fig~\ref{l=0figzeta1R}, the $\omega_R$
values for $\epsilon=1$(solid line) and $\epsilon=-1$(dashed line) are almost identical for small $Q_m$, a discernible difference emerges beyond approximately 0.4. Moreover, $\omega_R$
gradually increases with increasing $Q_m$, which indicates that the oscillation frequencies of the black hole perturbation become faster. Note that the QNM frequencies for $\epsilon=-1$ become slightly higher than those for $\epsilon=1$.
In addition, the absolute value of imaginary part $\lvert\omega_I\lvert$ monotonically decreases with increasing magnetic charge $Q_m$, see Fig.~\ref{l=0figzeta1I}. Physically, this signifies that the damping rate of the perturbation slows down due to $\omega_I<0$; that is, the lifetime of the quasinormal mode $(T\propto 1/\lvert\omega_I\lvert)$ becomes longer. 

The influence of parameter $\epsilon$ on the damping rate is very pronounced. Across the entire range of $Q_m$, the ``lengths" of the QNM frequency curves for $\epsilon=1$ and $\epsilon=-1$ are different. The reason is that the black hole with $\epsilon=-1$ becomes an extremal one more rapidly than others. On the other hand, $\lvert\omega_I\lvert$ for $\epsilon=-1$ is consistently smaller than that, comparing with $\lvert\omega_I\lvert$ with $\epsilon=1$. This means that within the theoretical framework of $\epsilon=-1$, the dilaton perturbation decays significantly more slowly, and the modes have a longer lifetime compared to the $\epsilon=1$ case. For instance, when $Q_m=0.6$, the lifetime for the $\epsilon=-1$ mode is approximately 9.43, indeed longer than the $\sim$9.26 for the $\epsilon=1$ mode from Table ~\ref{table-A-1}. 

In Fig.~\ref{l=0figzeta1I}, the curve  ($\epsilon=-1$) terminates before reaching the boundary of the parameter range due to the fact that the black hole arrives at extremal point 
($Q_m=0.826$) as the $Q_m$ increases. Here, it is worthy to note that the observed behavior of the imaginary part of the frequency (specifically, its tendency toward zero with increasing $Q_m$) does not occur for the $\epsilon=-1$ case. Hence, the curves terminate abruptly before this critical point to ensure the validity of the numerical method. Similar things will happen for all $\omega_I$ curves ($\epsilon=-1$) in Figs. 4, 5, and 6.  

Moreover, there are no positive imaginary of QNM frequencies for the dilaton mode with $l=0$. This indicates clearly  that this black hole is stable against the dilaton with $l=0$.

\subsection{ $l=1$}

In this case, we note that  two Zerilli variables  $R_1$ and $K$ are also  redundant.
For $l=1$ case, the polar linearized equation is given by the dilaton equation
\begin{eqnarray}
   \frac{d^{2}\Psi_d(r_{*})}{{dr_{*}}^{2}}+[\omega^{2}-V_{{\rm eff},l=1}(r)]\Psi_d(r_{*})=0,
   \label{wavefunc-2}
\end{eqnarray}
where the effective potential can be significantly simplified by employing the background field equations:
\begin{eqnarray}
V_{{\rm eff},l=1}(r) &&=\frac{B A'}{2 r}+\frac{A e^{-2 \phi}}{2 r^8}\Big[ r^6 e^{2 \phi }\left(r B'+8 r^2 B \phi '^2+4\right) \nonumber \\
&&+4 r^4 Q_{m}^2-12 \epsilon  Q_{m}^4 \left(e^{4 \phi }+1\right)\Big].
\end{eqnarray}
Before we proceed, we would like to mention the existing  QNM frequencies for $l=1$ scalar mode.
Even though Ref.~\cite{Bakopoulos:2024hah} showed the  QNM frequencies for a minimally coupled scalar, the results are not suitable for our comparison. 
Hence, we present the QNM frequencies (AIM) for the $l=1$ scalar mode~\cite{Zhang:2025xqt}
\begin{eqnarray}
&&M\omega \approx 0.297489-0.098155i(\epsilon=1), \quad M\omega \approx 0.297510-0.098099i(\epsilon=-1) \label{omegal103}
\end{eqnarray}
with  $M=1$ and $Q_m=0.3$. For $M=1$ and $Q_m=0.6$, these are 
\begin{eqnarray}
&&M\omega \approx0.313483-0.100351i(\epsilon=1), \quad M\omega \approx 0.312951-0.109943i(\epsilon=-1).  \label{omega106}
\end{eqnarray}
These are nearly the same as the dilaton with $l=1$ (AIM) in Table II. 

\begin{figure}[H]
\centering
\subfigure[~$\epsilon=1$]{
\label{figVeffl1zeta1} 
\includegraphics[height=1.5in]{
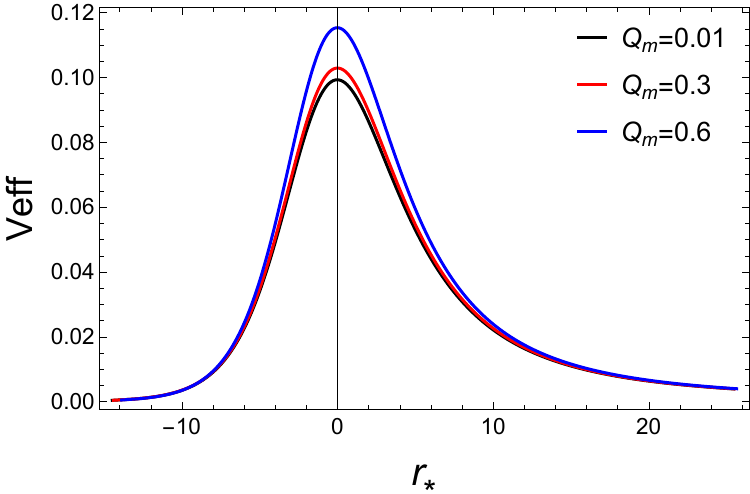}}
\quad
\subfigure[~$\epsilon=-1$]{
\label{figVeffl1zetam1} 
\includegraphics[height=1.5in]{
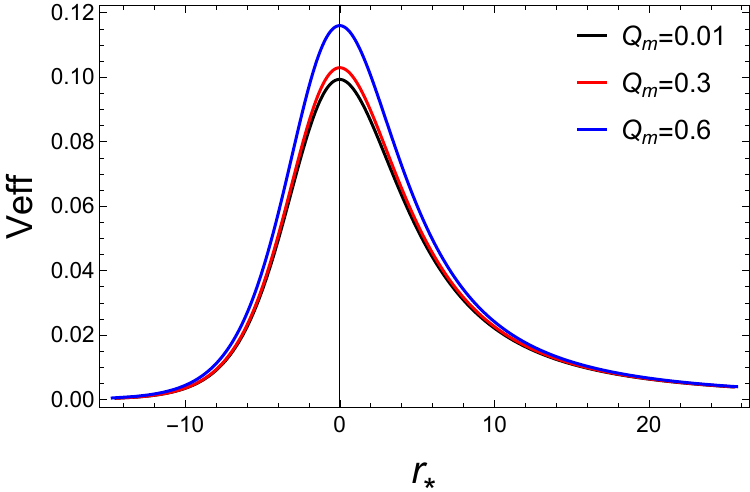}}
\caption{
Graphs of the potential for  the $l=1$ dilaton as function of $r_*$.
}
\label{fig:l=1Vall}
\end{figure}

As an evidence  of positive definite potentials  is shown in Fig. \ref{fig:l=1Vall}, 
we found from Table \ref{table-B-1} that there are no positive imaginary part of QNM frequencies for the $l=1$  dilaton. This indicates that this mode is stable against the dilaton perturbation.

\begin{table*}[h!]
\caption{Fundamental ($n=0$) QNM frequencies for $l=1$-mode  dilaton  $\Psi_d$ and coupling parameters $\epsilon=\pm1$ around dEH black holes with $M=1$,  It indicates  values from the DI and AIM  across different $Q_m$ and the discrepancy between them.}
\label{table-B-1}
\resizebox{\linewidth}{!}{
\begin{tabular}{|c|c|c|c|c|c|c|c|} \hline
 & & \multicolumn{3}{|c|}{$\epsilon=1$}  &  \multicolumn{3}{|c|}{$\epsilon=-1$} \\ \hline
$l$ &  	$Q_m$ & DI  &  AIM  & $\Delta_{DA}$   &  DI  &  AIM   & $\Delta_{DA}$ \\ \hline
 \multirow{3}{*}{1}&  0.01  &   $0.292999 -0.0977016 i$  &  $0.293098-0.0976759 i$   &0.0223437\%   & $0.292999 -0.0977016 i$        & $0.293098-0.0976758 i$  & 0.0223443\%  \\ 
& 0.3 &  $0.298884-0.0983378 i$  & $0.298820-0.0983221 i$    &0.0207687\%   & $0.298538-0.0982363 i$   &  $0.298876-0.0982566 i$  &   $0.0206404\%$   \\ 
&  0.6 &  $0.318099-0.101163 i$ &$0.318074-0.101219 i$   &  $0.018361\%$ &$0.319515-0.0993235 i$  & $0.319499-0.0993733 i$ & $0.0156504\%$ \\ \hline
\end{tabular}}
\end{table*}

\begin{figure}[H]
\centering
\subfigure[~$\omega_{R}$ vs $ Q_m$]{
\label{l=1figzeta1R} 
\includegraphics[height=1.5in]{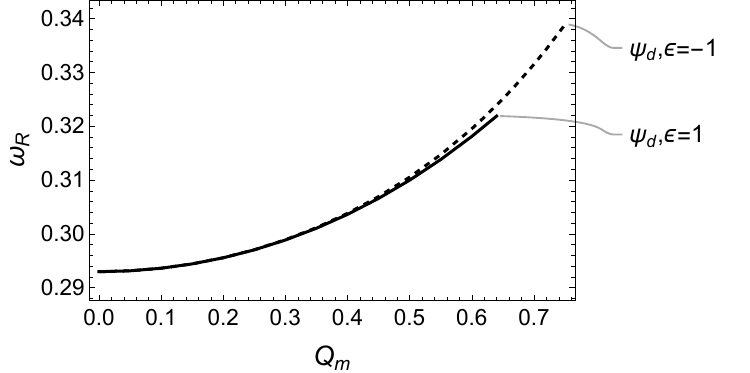}}
\quad
\subfigure[~$\omega_{I}$ vs $ Q_m$]{
\label{l=1figzeta1I} 
\includegraphics[height=1.5in]{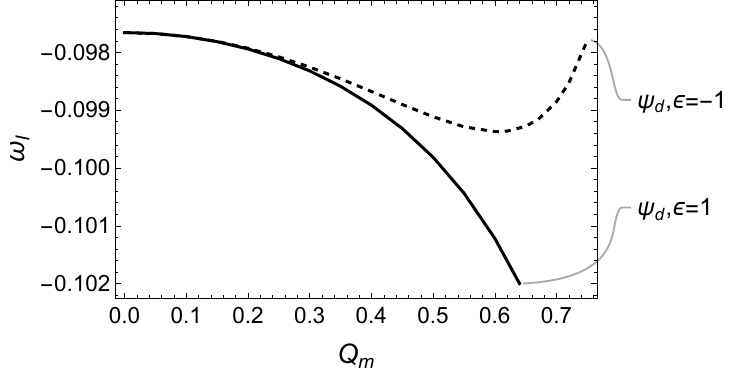}}
\caption{Variation of fundamental QNM frequencies (real and  imaginary parts) with $Q_m$ for the $l=1$ dilaton $\Psi_d$ around dEH black holes with $M=1$. Solid lines indicate $\epsilon=1$ and  dashed lines are for $\epsilon=-1$.}
\label{fig:l=1all}
\end{figure}

It is clear that the influences of parameters $Q_m$ and $\epsilon$ on the QNM frequencies of $l=1$ are similar to those of $l=0$.
Fig. \ref{fig:l=1all} shows that  $\omega_R$ of $l=1$ mode $\Psi_d$ increases  as $Q_m$ increases and its $\epsilon=\pm1$ behavior shows  slight  difference  as $Q_m$ increases. $\omega_I$ of $l=1$ mode decreases   as $Q_m$ increases and its $\epsilon=\pm1$ behavior indicates  a  difference  as $Q_m$ increases.

\subsection{$l\geq2$}
In this case, 
we can obtain the fundamental QNM frequencies of both metric and dilaton perturbations. The eigen-frequencies were obtained for different quantum number  $l=2,3$ with $\epsilon=1$ and $\epsilon=-1$. The fundamental QNM frequencies for the metric and  dilaton perturbations under different  $Q_m$ are listed in Table \ref{table-1} and \ref{table-2}, respectively.  In the two tables, we present the QNM  frequencies obtained via the direct integration and continued fraction methods. The relative discrepancy $(\Delta_{DC}) $ is predominantly below $1\%$, and for very small $Q_m$ values, it can be as low as approximately $0.01\%$. This high degree of consistency firmly validates the reliability and precision of the numerical approach, providing a solid foundation for subsequent physical interpretation.

For both cases of $\epsilon=\pm1$ with $Q_m=0$, the metric function in Eq. \eqref{metric} reduces to the Schwarzschild solution.
First of  all, we mention that the Schwarzschild spacetime is stable and its fundamental QNM frequency ($\omega=M\omega_R+M\omega_I$)
for the $l=2$ case is listed~\cite{Pani:2013,Berti:2009kk, Blazquez-Salcedo:2016enn, Chandrasekhar:1975zza}

\begin{eqnarray}
&&M\omega \approx 0.3737-0.08896i, \quad {\rm gravitational~ mode} \nonumber \\
&&M\omega \approx 0.4836-0.09676i, \quad {\rm scalar~ mode}. \label{omegal2}
\end{eqnarray}
For $l=3$ mode, these read as
\begin{eqnarray}
&&M\omega \approx 0.599443-0.0927025i, \quad {\rm gravitational ~mode} \nonumber \\
&&M\omega \approx 0.675747-0.0965941i, \quad {\rm scalar ~ mode}. \label{omegal3}
\end{eqnarray}
The negative $\omega_I$ implies that the Schwarzschild black hole is stable against metric and scalar perturbations.
For $Q_m=0.01$, the QNM  frequencies in Table II and III exhibit a slight deviation from those of  Schwarzschild case.

Although the dEH black hole  with  $\epsilon=-1$  have  two horizons, this frequency shift is still observed.  Both Tables list the QNM frequencies for gravitational ($l=2,3$) and dilaton  ($l=2,3$)  perturbations across two  $\epsilon=\pm1$ and three  $Q_m=0.01,0.3,0.6$ obtained by the direct integration and continued fraction methods.  It  reveals a strong consistency between the two computational approaches. All negative $\omega_I$ imply that the dEH black hole is stable against polar metric and dilaton perturbations with $l=2,3$.

\begin{table}[h!]
\caption{Fundamental  QNM frequencies  for $l=2,3$ gravitational modes $\Psi_g$  and coupling parameters $\epsilon=\pm1$ around  dEH black hole with $M=1$, It shows values from  DI and continued fraction method (CFM)  across different $Q_m$ and the discrepancy between them.}
\label{table-1}
\resizebox{\linewidth}{!}{
\begin{tabular}{|c|c|c|c|c|c|c|c|} \hline
 &  & \multicolumn{3}{|c|}{$\epsilon=1$}  &  \multicolumn{3}{|c|}{$\epsilon=-1$} \\ \hline
$l$ &  	$Q_m$ & DI  &  CFM  & $\Delta_{DC}$   &  DI  &  CFM   & $\Delta_{DC}$ \\ \hline
 \multirow{3}{*}{2}&  0.01  &   $0.373681\, -0.0889637 i$  &  $0.373682\, -0.0889625 i$   &0.000341888\%   & $0.373681\, -0.0889637 i$        & $0.373682\, -0.0889625 i$  & 0.000341832\%  \\
& 0.3 &  $0.383315\, -0.0897244 i$  & $0.381701\, -0.0898252 i $    &0.411593\%   & $0.384062\, -0.0895549 i$   &  $0.381902\, -0.0896742 i$  &   $0.550222\%$   \\
&  0.6 &  $0.407109\, -0.0940574 i$ &$0.407641\, -0.0937156 i$   &  $0.151336\%$ &$0.425836\, -0.0877879 i$  & $0.411569\, -0.0911675 i$ & $3.42445\%$ \\ \hline

 \multirow{3}{*}{3}&  0.01  &   $0.599455\, -0.0927048 i$  &  $0.599455\, -0.0927036 i$   &0.000200698\%   & $0.599455\, -0.0927048 i$        & $0.599455\, -0.0927036 i$  & 0.000720091\% \\
& 0.3 &  $0.610899\, -0.0928186 i$  & $0.610231\, -0.0933314 i$    &0.136312\%   & $0.61059\, -0.0939475 i$   &  $0.61126\, -0.0925498 i$  &   $0.205123\%$  \\
&  0.6 &  $0.645185\, -0.0961996 i$ &$0.647846\, -0.0968344 i$   &  $0.418561\%$ &$0.654047\, -0.0890607 i$  & $0.643508\, -0.0949577 i$ & $1.84296\%$ \\  \hline
\end{tabular}}
\end{table}

\begin{table*}[h!]
\caption{Fundamental  QNM frequencies for $l=2,3$  dilaton modes  $\Psi_d$ and coupling parameters $\epsilon=\pm1$ around the  dEH black holes with $M=1$,  It indicates  values from DI and CFM  across different $Q_m$ and the discrepancy between them.}
\label{table-2}
\resizebox{\linewidth}{!}{
\begin{tabular}{|c|c|c|c|c|c|c|c|} \hline
 & & \multicolumn{3}{|c|}{$\epsilon=1$}  &  \multicolumn{3}{|c|}{$\epsilon=-1$} \\ \hline
$l$ &  	$Q_m$ & DI  &  CFM  & $\Delta_{DC}$   &  DI  &  CFM   & $\Delta_{DC}$ \\ \hline
 \multirow{3}{*}{2}&  0.01  &   $0.483652\, -0.0967603 i$  &  $0.483644\, -0.0967588 i$   &0.00161798\%   & $0.483652\, -0.0967603 i$        & $0.483653\, -0.0967594 i$  & 0.000274627\%  \\ 
& 0.3 &  $0.491923\, -0.0973135 i$  & $0.491903\, -0.0973215 i$    &0.00423365\%   & $0.491973\, -0.0972586 i$   &  $0.491954\, -0.097266 i$  &   $0.00408258\%$   \\ 
&  0.6 &  $0.5195\, -0.0995496 i$ &$0.519121\, -0.0997137 i$   &  $0.0780621\%$ &$0.520771\, -0.0980041 i$  & $0.520439\, -0.0981907 i$ & $0.0718172\%$ \\ \hline

 \multirow{3}{*}{3}&  0.01  &   $0.675378\, -0.0964999 i$  &  $0.675378\, -0.0965002 i$   &0.0000404289\%   & $0.675378\, -0.0964999 i$        & $0.675378\, -0.0965002 i$  & 0.0000404331\%  \\ 
& 0.3 &  $0.686337\, -0.0970269 i$  & $0.686329\, -0.0970358 i$    &0.00177566\%   & $0.686396\, -0.0969773 i$   &  $0.686388\, -0.0969857 i$  &   $0.00167966\%$   \\ 
&  0.6 &  $0.722614\, -0.0991309 i$ &$0.72245\, -0.0992895 i$   &  $0.0312943\%$ &$0.724121\, -0.0977703 i$  & $0.724006\, -0.0979284 i$ & $0.0267559\%$ \\ \hline
\end{tabular}}
\end{table*}

\begin{figure}[H]
\centering
\subfigure[~$\omega_{R}$ vs $Q_m$]{
\label{figzeta1R} 
\includegraphics[height=1.7in]{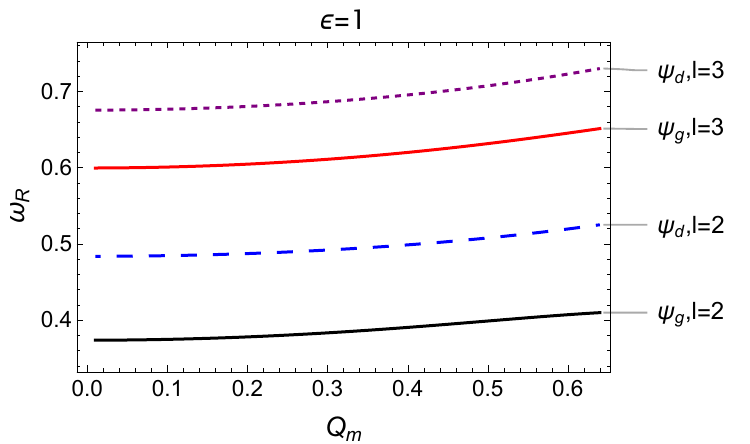}}
\quad
\subfigure[~$\omega_{I}$ vs $ Q_m$]{
\label{figzeta1I} 
\includegraphics[height=1.7in]{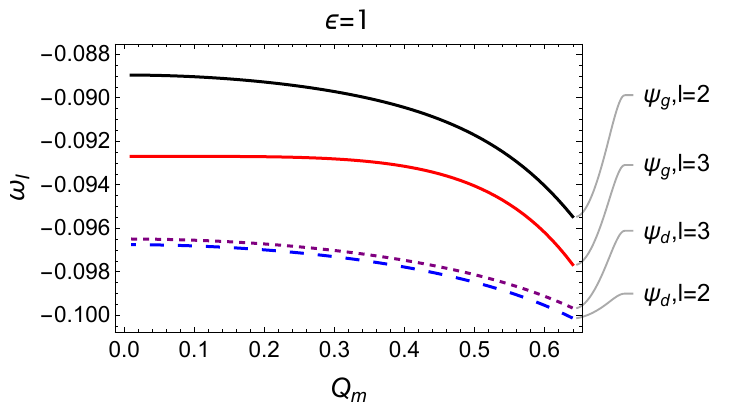}}
\caption{Variation of real and imaginary parts for fundamental QNM frequencies with $Q_m$ for gravitational ($\Psi_g$: solid lines) and dilaton ($\Psi_d$: dashed lines)   around   dEH black holes with $M=1$ and  $\epsilon=1$. Black and red indicate  gravitational mode with $l=2,3$ while  blue and purple show  dilaton mode with  $l=2,3$.}
\label{fig:all1}
\end{figure}

\begin{figure}[H]
\centering
\subfigure[~$\omega_{R} $ vs $ Q_m$]{
\label{figzeta-1R} 
\includegraphics[height=1.7in]{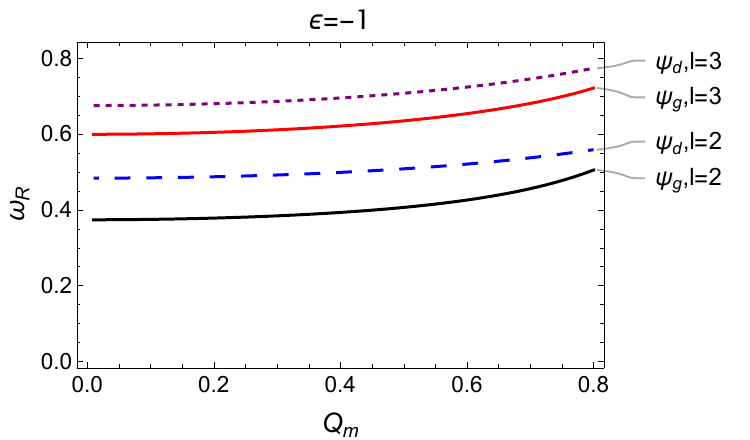}}
\quad
\subfigure[~$\omega_{I} $ vs $ Q_m$]{
\label{figzeta-1I} 
\includegraphics[height=1.7in]{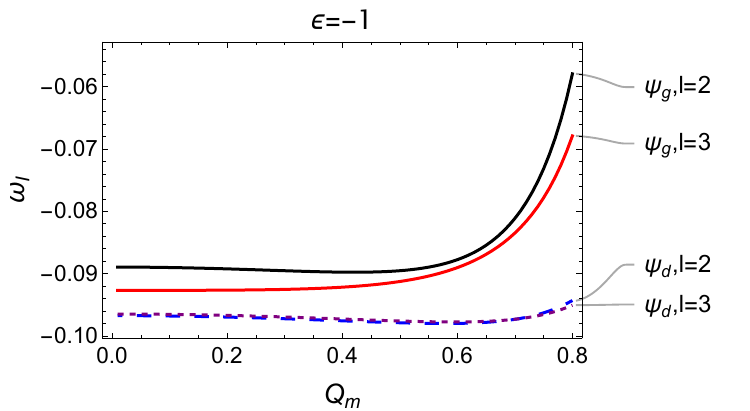}}
\caption{Variation of real and imaginary parts for fundamental QNM frequencies with $Q_m$ for gravitational ($\Psi_g$: solid lines) and dilaton  ($\Psi_d$: dashed lines) modes  around  $M=1$ dEH black holes at coupling $\epsilon=-1$ .  Black and red denote for gravitational mode with  $l=2,3$ whereas blue and purple represent  dilaton mode with  $l=2,3$.}
\label{fig:all2}
\end{figure}

Concerning interpretation of  QNM of the dEH black holes, the real part of the frequencies governs the oscillation rate, and the imaginary part indicates  the decay rate (for a negative value,  it means  damping).  Figs. \ref{fig:all1} and  \ref{fig:all2} depict the variation with $Q_m$ of the real and imaginary parts of the fundamental QNM frequencies, computed from  the direct integration method for $\epsilon=\pm1$.

As shown in Fig. \ref{figzeta1R} for the case of $\epsilon=1$, the real parts of the QNM frequencies for both the gravitational mode $\Psi_g$ and the dilaton mode $\Psi_d$ with $l=2,3$ increase monotonically with increasing magnetic charge $Q_m$. This indicates that an increase in the magnetic charge $Q_m$ leads to a higher oscillation frequency. As illustrated in Fig. \ref{figzeta1I} for the imaginary part, the QNM frequencies of the gravitational mode and dilaton mode decrease with increasing magnetic charge $Q_m$, implying that for this case, an increase in the magnetic charge $Q_m$ results in a faster damping rate.

For the case of $\epsilon=-1$, the real part (Fig.\ref{figzeta-1R}) exhibits the same trend as that for $\epsilon=1$: all modes increase monotonically with $Q_m$. The imaginary part (Fig.\ref{figzeta-1I}), with the exception of the mode $l=3$ (which increases monotonically), undergoes an initial decrease followed by an increase, with a rapid growth before the extremal charge $Q_m\sim 0.826$. Taking into account the validity of the numerical method, the curves end before reaching $Q_m=0.826$. Furthermore, for fixed $Q_m$ and $\epsilon$, the oscillation frequencies $(\omega_R)$ for the $l=3$ modes are systematically higher than their $l=2$ counterparts, consistent with the expectation that higher-order multipole perturbations correspond to higher characteristic frequencies.

\section{Conclusion and discussion}
\label{sec6}

In this paper, we have obtained  the QNM frequencies  of polar gravitational and dilaton  perturbations around the dEH black holes within the dilaton-Einstein-Maxwell theory.
This theory is obtained by combining Einstein-Maxwell-dilaton with  a dilaton coupling to EH term.   Here, the EH term serves  as a nonlinear extension of QED, providing the magnetically charged EH black hole solution~\cite{Yajima:2000kw}. 

The perturbation theory of metric and dilaton  around  the dEH  black hole described by \eqref{metric} and \eqref{solution} is necessary to compute the QNM frequencies and thus, to study the stability of the dEH black holes.  The polar part of this perturbation  theory  became  four coupled equations (\ref{eq:Kp})-(\ref{eq:phi1pp}). To obtain the QNM frequencies of this system, we have employed  the  direct integration and  matrix-valued continued fraction methods. We have shown that  the results obtained from these two numerical techniques are in very good agreement.  The fundamental QNM frequencies for dilaton and metric  perturbations with $\epsilon=\pm 1$ are summarized in Tables \ref{table-A-1}-\ref{table-2}. Their dependence on the magnetic charge $Q_m$ is depicted in Figs. \ref{fig:l=0all}, \ref{fig:l=1all}, \ref{fig:all1}, and \ref{fig:all2}.  A notable feature is  that the distinct behavior appeared in the imaginary part of the QNM frequencies for  $\epsilon=-1$ configuration.
Consequently, all negative imaginary parts of QNM frequencies imply  that the dEH black hole  with dilaton hair is stable against polar-metric with $l=2,3$ and dilaton with $l=0,1,2,3$.

Moreover, we expect that  observed deviations in the QNM frequencies from general relativity predictions may  serve as an observational signature for constraining or detecting the black hole's magnetic charge $Q_m$ and coupling strength $\epsilon$ in future gravitational-wave astronomy.
This work has been restricted to obtaining  the QNM frequencies for polar perturbations. Exploring other possibility of axial  perturbations together with  electromagnetic perturbation will  make  the stability of the dEH black holes clear.

 \vspace{1cm}
{\bf Acknowledgments}
 \vspace{1cm}

We gratefully acknowledge support by the National Natural Science Foundation of China (NNSFC) (Grant No.12365009).  Y.S.M. was supported by the National Research Foundation of Korea (NRF) grant funded by the Korea government(MSIT) (RS-2022-NR069013).

\appendix
\section{Zerilli  and dilaton equations}
\label{appendix-A}
First of all, Eqs. \eqref{eq:Kp}-\eqref{eq:constraint} have been simplified using the background solution. Specifically, the exponential factors of $\bar{\phi}(r)$ and their second derivatives have been eliminated, and a conversion between derivatives of $A(r)$ and $B(r)$ has been performed. Making use of  Eq.\eqref{eq:constraint}, we may  eliminate $H_0(r)$ from the system.  It is convenient to use the redefinition
\begin{eqnarray}
H_{1}(r)=\omega R_{1}(r).\label{R1def}
\end{eqnarray}
Now, we are left with two first-order equations for $K(r)$ and $R_1(r)$, but   they can be transformed  to the  single second-order 
equation. As was shown in ~\cite{Zerilli:1970se} and \cite{Zerilli:1974ai}, we adopt the procedure of Ref. \cite{Brito:2018hjh} and acknowledge the authors for using their \texttt{Mathematica} notebook  on GitHub.  
We introduce $\hat{K}(r)$ and $\hat{R}(r)$ as 
\begin{eqnarray}
&&K(r)=\alpha(r)\hat{K}(r)+\beta(r)\hat{R}(r),\label{Khatdef}\\
&&R_1(r)=\gamma(r)\hat{K}(r)+\lambda(r)\hat{R}(r).\label{Rhatdef}
\end{eqnarray}
Substituting Eqs.\eqref{Khatdef} and \eqref{Rhatdef} into the system, we find  
\begin{eqnarray}
&&\frac{d\hat{K}(r)}{dr_*}=\hat{R}(r)+{\rm matter\,couplings},\label{eq:Khat}\\
&&\frac{d\hat{R}(r)}{dr_*}+\omega^2\hat{K}(r)=V_{K}(r)\hat{K}(r)+{\rm matter\,couplings},\label{eq:Rhat}
\end{eqnarray}
where $V_K(r)$ is a potential and the matter couplings denotes  $\delta\phi_1(r)$ and its derivative.   By taking the derivative of Eq.\eqref{eq:Khat} with respect to the tortoise coordinate $r_*$ together with Eq.\eqref{eq:Rhat}, we obtain 
\begin{eqnarray}
&&\Big[\frac{d^2}{dr_{*}^{2}}+\omega^2\Big]\hat{K}(r)=V_{K}\hat{K}(r)+{\rm matter\,couplings}.
\label{eq:Zerilli-2}
\end{eqnarray}
Here,  we obtain
\begin{eqnarray}
\alpha(r)=&&-\Big\{A Q_{m}^2 \left(4 A^3-4 A^2 \left(B+l^2+l\right)+A \left(2 B \left(-r A'+l^2+l\right)+l^2 (l+1)^2\right)\right.\nonumber\\
&&\left.+B r \left(2 B r A''+l (l+1) A'\right)\right)\Big\}/\Big\{2 r^3 \sqrt{A B} \bar{\phi}' \left(-2 A^2+B r A'+A l (l+1)\right)\Big\},\label{eq:alpha}\\
\beta(r)=&&-\frac{Q_m^2}{r^2 \bar{\phi} '},\label{eq:beta}\\
\gamma(r)=&&-\frac{i Q_m^2 \left(4 A^3+B r^2 \left(A'\right)^2+A r \left(l (l+1) A'-2 B r A''\right)-2 A^2 \left(r A'+l^2+l\right)\right)}{2 r^2 \sqrt{A B} \bar{\phi}' \left(-2 A^2+B r A'+A l (l+1)\right)},\label{eq:gamma}\\
\lambda(r)=&&\frac{i Q_m^2}{B r \bar{\phi}'}.\label{eq:lambda}
\end{eqnarray}

But, the derivation of  dilaton  equation is straightforward. In Eq.\eqref{eq:phi1pp},  $\delta\phi_1(r)$ is related to $\hat{S}(r)$ as  $\delta\phi_1(r)=\hat{S}(r)$. By combining  $K$ from Eq.\eqref{Khatdef} with  $\hat{R}$ obtained by inverting Eq.\eqref{eq:Khat}, Eq.\eqref{eq:phi1pp} leads to  Eq.\eqref{eq:Scalar}.

\section{Asymptotic Iteration Method}
\label{appendix-AIM}
In this appendix, we mention briefly the asymptotic iteration method (AIM). 
Firstly, we express  dilaton equation \eqref{wavefunc-1} for $l=0$ mode in terms of $u=1-r_h/r$ as 
\begin{eqnarray}
  &&\psi''(u)+\frac{1}{2} \left( \frac{4}{u-1 } + \frac{A'(u)}{A(u)} + \frac{B'(u)}{B(u)} \right) \psi'(u) 
+\frac{1}{2r_h^6(u-1)^4A(u)B(u)} 
\Bigg[2r_h^8\omega^2\nonumber\\
&&+(u-1)^3\Bigg(r_h^6B(u)A'(u)\Big(1 + 3 (u-1)^2 (\phi'(u))^2\Big)
+A(u)\Big(-2 e^{-2\phi(u)} r_h^4 (u-1)Q_{m}^2\nonumber\\
&&\times  \Big( 2 + (u-1) \phi'(u) \Big)
+12 (u-1)^5 \epsilon Q_{m}^4\Big( 2 \cosh(2\phi(u)) - (u-1) \sinh(2\phi(u)) \phi'(u) \Big)+r_h^6 B'(u)\nonumber\\
&&+r_h^6\Big((u-1)
\Big(4B(u)-(u-1)B'(u)\Big)(\phi'(u))^2 -2 (u-1)^3 B(u)(\phi'(u))^4\Big)\Big)\Bigg)\Bigg]\psi(u)=0.\label{equ1}
\end{eqnarray}
Here, the range of $u$  is given by  $0\leqslant u<1$. 

From \eq{equ1}, we examine  the behavior of the function $\psi(u)$ at horizon $(u=0)$ and at the boundary $u=1$. Near the horizon $(u=0)$, we have $A(0)\approx u A'(0)$ and $B(0)\approx u B'(0)$. Then, \eq{equ1} becomes
\begin{eqnarray}
  \psi''(u)+\frac{1}{u}\psi'(u)+\frac{r_h^2\omega^2}{u^2 A'(0) B'(0)}\psi(u)=0
\end{eqnarray}
whose solution is given by
\begin{eqnarray}
  \psi(u\to 0)\sim C_1 u^{-\xi}+C_2 u^{\xi},\quad \xi=\frac{ir_h\omega}{\sqrt{A'(0) B'(0)}}.
\end{eqnarray}
In this case,  we  set $C_2=0$  to respect the ingoing condition at the horizon.
At infinity $(u=1)$, the asymptotic form of \eq{equ1} takes the form 
\begin{eqnarray}
  \psi''(u)-\frac{2}{1-u}\psi'(u)+\frac{r_h^2\omega^2}{(1-u)^4}\psi(u)=0
\end{eqnarray}
with $A(1)=1$ and $B(1)=1$.
Then, we obtain the solution
\begin{eqnarray}
  \psi(u\to 1)\sim D_1 e^{-\zeta}+D_2 e^{\zeta},\quad  \zeta=\frac{ir_h\omega}{1-u}.
\end{eqnarray}
To implement the outgoing boundary condition, we choose  $D_1=0$.

Using the  solutions at horizon and infinity, we may  define the ansatz for $\psi(u)$  as
\begin{eqnarray}
  \psi(u)=u^{-\xi}e^{\zeta}\chi(u) \label{eqschi}.
\end{eqnarray}
Substituting  \eq{eqschi} to \eq{equ1} leads to 
\begin{eqnarray}
  \chi''=\lambda_0(u)\chi'+s_0(u)\chi \label{eqschieq},
\end{eqnarray}
where 
\begin{eqnarray}
  \lambda_0(u)=\frac{1}{2} \left(\frac{4 i r_h \omega }{u \sqrt{A'(0)} \sqrt{B'(0)}}-\frac{A'(u)}{A(u)}-\frac{B'(u)}{B(u)}-\frac{4 \left(i r_h\omega +u-1\right)}{(u-1)^2}\right),
\end{eqnarray}
and
\begin{eqnarray}
  s_0(u)&=&\frac{A'(u)}{2A(u)(1-u)}+r_h\omega\Big(\dfrac{r_h \omega}{(u-1)^4} + \dfrac{r_h \omega}{u^2 A'(0) B'(0)} + \dfrac{i (u^2 - 1 + 2i r_h u \omega)}{(u-1)^2 u^2 \sqrt{A'(0)} \sqrt{B'(0)}}\Big) \nonumber\\
  &&-2 \left( \phi'(u) \right)^2 +  (u-1)^2 \left( \phi'(u) \right)^4-\frac{r_h^2\omega^2}{A(u)B(u)(u-1)^4}-\frac{A'(u)}{2A(u)}\Big(3(u-1)(\phi'(u))^2\Big) \nonumber\\
  &&-\frac{i r_h \omega A'(u)}{2A(u)}\Big(  \dfrac{1}{(u-1)^2} - \dfrac{1}{u \sqrt{A'(0)} \sqrt{B'(0)}}\Big)+\dfrac{Q_m^2}{r_h^2B(u)}\Big(e^{-2\phi(u)}  \left( 2 + (u-1) \phi'(u) \right)\Big) \nonumber\\
 &&+ \dfrac{1}{r_h^6B(u)}6Q_m^4(u-1)^4 \epsilon  \Big(-2 \cosh(2\phi(u)) + (u-1) \sinh(2\phi(u)) \phi'(u) \Big)\nonumber\\
 &&+\frac{B'(u)}{2B(u)}\Big(-\frac{-1 + u + i r_h \omega}{(u-1)^2} 
+ \frac{i r_h\omega}{u \sqrt{A'(0)} \sqrt{B'(0)}} 
+ ( u-1) (\phi'(u)^2\Big).
  \end{eqnarray}
Based on $\lambda_0(u)$ and $s_0(u)$, Eq.\eqref{eqschieq} can be solved numerically by using the  AIM \cite{Cho:2009cj}. 

Similarly, we  obtain functions  $\lambda_0$ and  $s_0$  for $l=1$ case of Eq.\eqref{wavefunc-2} as   
\begin{eqnarray}
  \lambda_0(u)=\frac{1}{2} \left(\frac{4 i r_h \omega }{u \sqrt{A'(0)} \sqrt{B'(0)}}-\frac{A'(u)}{A(u)}-\frac{B'(u)}{B(u)}-\frac{4 \left(i r_h \omega +u-1\right)}{(u-1)^2}\right),
\end{eqnarray}
and
\begin{eqnarray}
  s_0(u)&=&\frac{r_h^{2}\omega^{2}}{(u-1)^{4}} 
- \frac{(u-1+ir_h\omega)A'(u)}{2(u-1)^{2}A(u)} 
+ \frac{r_h^{2}\omega^{2}}{u^{2}A'(0)B'(0)}+8(\phi'(u)^2\nonumber\\
&&+\frac{ir_h\omega 
}
{2(u-1)^{2}u^{2}A(u)\sqrt{A'(0)}\sqrt{B'(0)}}\Big(2(u^{2}-1+2ir_h \omega \,u)A(u) 
+ (u-1)^{2}u A'(u)\Big)\nonumber\\
&&+\frac{1}{2B(u)}\Big(\frac{4}{(u-1)^2} 
- \frac{2r_h^2\omega^2}{(u-1)^4 A(u)} 
+ \frac{4e^{-2\phi(u)}Q_{m}^2}{r_h^2} 
- \frac{24(u-1)^4\epsilon\cosh(2\phi(u))Q_{m}^4}{rh^6}\Big)\nonumber\\
&&+\frac{B'(u)}{2B(u)}\Big(-\frac{-1 + u + i\,r_h\,\omega}{(-1+u)^2} 
+ \frac{i\,r_h\,\omega}{u\sqrt{A'(0)}\sqrt{B'(0)}}\Big).
\end{eqnarray}

\section{Convergence analysis for the continued fraction method}
\label{appendix-CFM}

\begin{figure}[H]
\centering
\subfigure[$\epsilon=1,Q_m=0.01$]{
\label{appendix-CFM-fig1a}
\includegraphics[height=1.2in]{
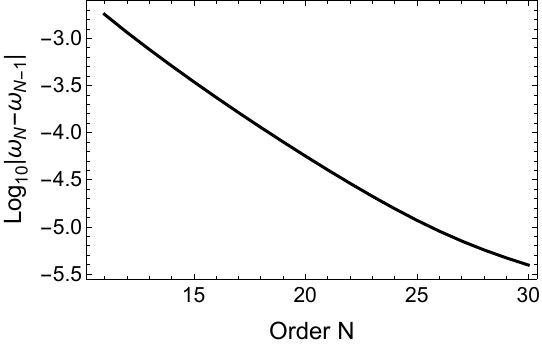}}
\quad
\subfigure[$\epsilon=1,Q_m=0.3$]{
\label{appendix-CFM-fig1b}
\includegraphics[height=1.2in]{
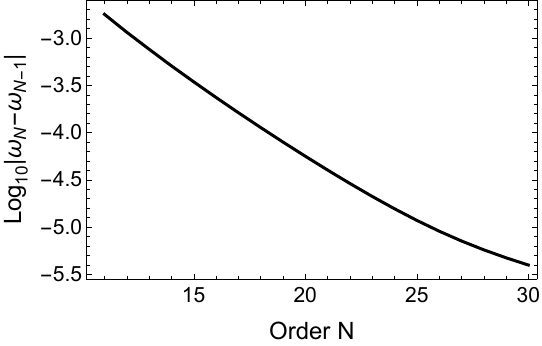}}
\quad
\subfigure[$\epsilon=1,Q_m=0.6$]{
\label{appendix-CFM-fig1c}
\includegraphics[height=1.2in]{
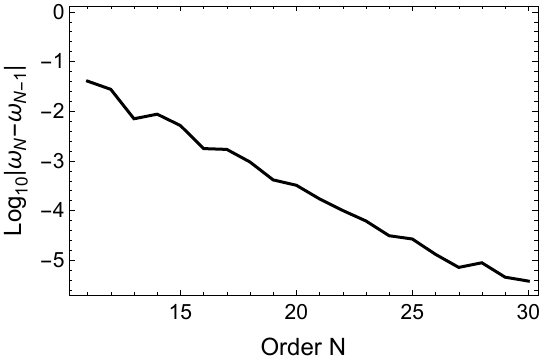}}
\quad
\subfigure[$\epsilon=-1,Q_m=0.01$]{
\label{appendix-CFM-fig1d}
\includegraphics[height=1.2in]{
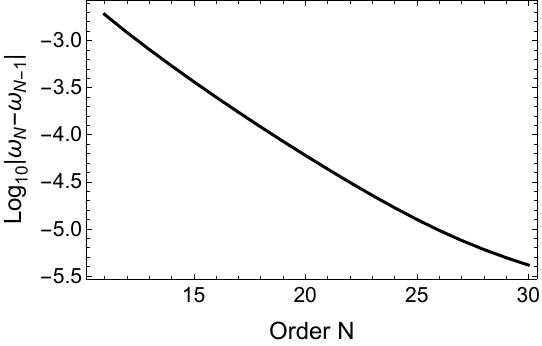}}
\quad
\subfigure[$\epsilon=-1,Q_m=0.3$]{
\label{appendix-CFM-fig1e}
\includegraphics[height=1.2in]{
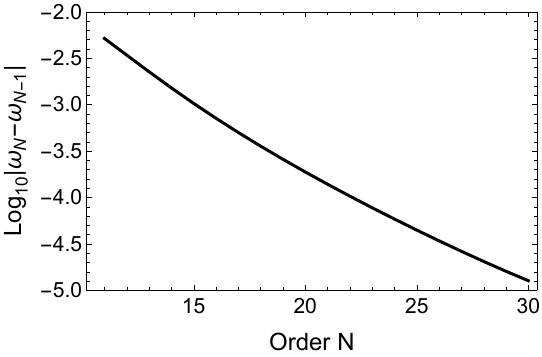}}
\quad
\subfigure[$\epsilon=-1,Q_m=0.6$]{
\label{appendix-CFM-fig1f}
\includegraphics[height=1.2in]{
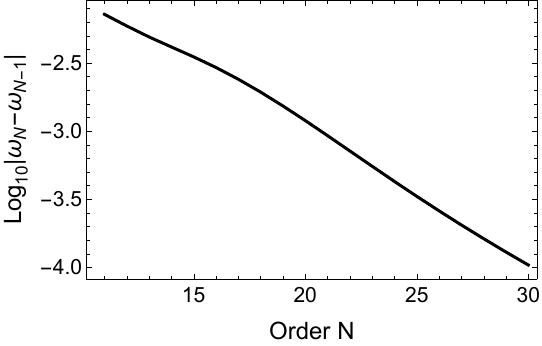}}
\caption{Variation of the successive differences for the QNM frequencies $\omega$ with respect to the order $N$ in CFM. This was done  for the gravitational $l=2$ mode ($\Psi_g$) with $\epsilon=\pm1$ and  $Q_m=0.01, 0.3, 0.6$. }
\label{appendix-CFM-fig1}
\end{figure}

\begin{figure}[H]
\centering
\subfigure[$\epsilon=1,Q_m=0.01$]{
\label{appendix-CFM-fig2a}
\includegraphics[height=1.2in]{
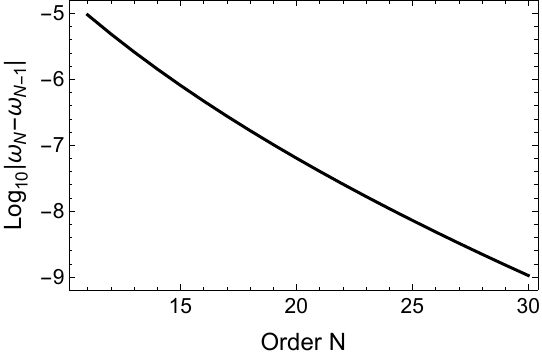}}
\quad
\subfigure[$\epsilon=1,Q_m=0.3$]{
\label{appendix-CFM-fig2b}
\includegraphics[height=1.2in]{
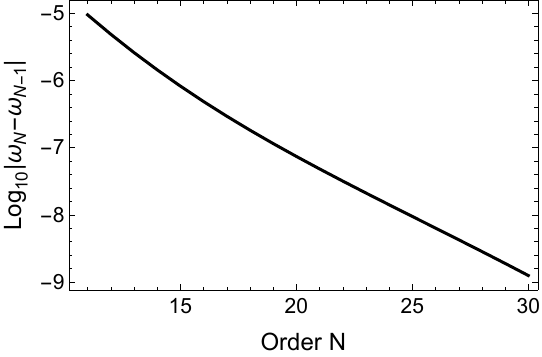}}
\quad
\subfigure[$\epsilon=1,Q_m=0.6$]{
\label{appendix-CFM-fig2c}
\includegraphics[height=1.2in]{
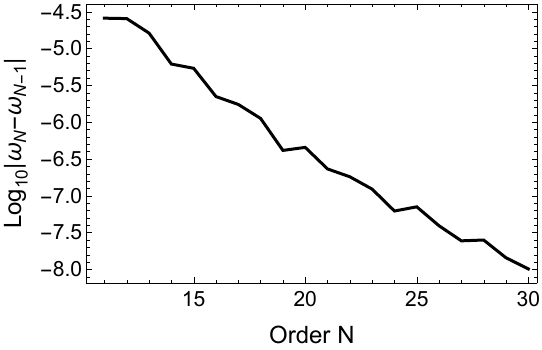}}
\quad
\subfigure[$\epsilon=-1,Q_m=0.01$]{
\label{appendix-CFM-fig2d}
\includegraphics[height=1.2in]{
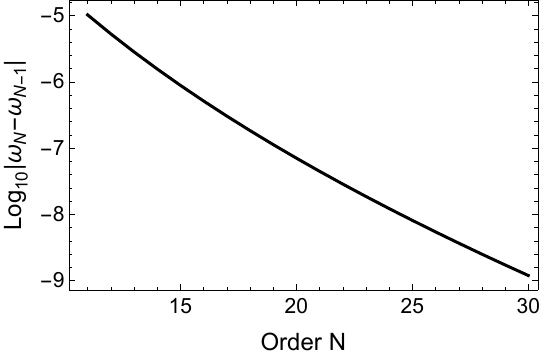}}
\quad
\subfigure[$\epsilon=-1,Q_m=0.3$]{
\label{appendix-CFM-fig2e}
\includegraphics[height=1.2in]{
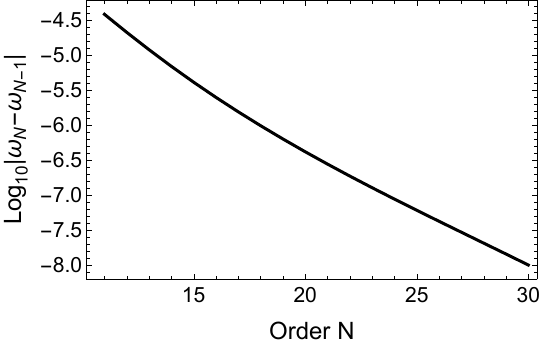}}
\quad
\subfigure[$\epsilon=-1,Q_m=0.6$]{
\label{appendix-CFM-fig2f}
\includegraphics[height=1.2in]{
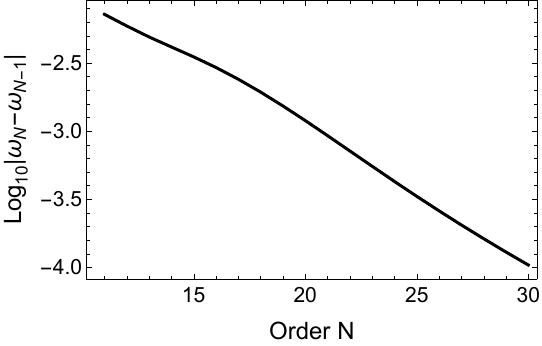}}
\caption{Variation of the successive differences for the QNM frequencies $\omega$ with respect to the order $N$ in CFM. This was done  for the $l=2$ dilaton ($\Psi_d$) with $\epsilon=\pm1$ and  $Q_m=0.01, 0.3, 0.6$.}
\label{appendix-CFM-fig2}
\end{figure}

In this appendix, we perform  convergence analysis for the continued fraction method (CFM).
As is shown in Figs.~\ref{appendix-CFM-fig1} ($\Psi_g$) and \ref{appendix-CFM-fig2} ($\Psi_d$), variation of the successive differences for the QNM frequencies $\log_{10}|\omega_N-\omega_{N-1}|$ decreases monotonically  as the order $N$ increases for $Q_m=0.01,0.3$, but it does not decrease smoothly for $\epsilon=1, Q_m=0.6$. The  latter curve displays minor oscillations superimposed on a clear descending trend, a feature commonly observed in numerical calculations and fully acceptable for the reliability of the results. This demonstrates  convergence with respect to numerical resolution for CFM data.

\end{document}